\documentclass[journal]{IEEEtran}
\usepackage{amsmath,amsfonts}
\usepackage{algorithmic}
\usepackage{array}
\usepackage[caption=false,font=normalsize,labelfont=sf,textfont=sf]{subfig}
\usepackage{textcomp}
\usepackage{stfloats}
\usepackage{url}
\usepackage{hyperref}
\usepackage{verbatim}
\usepackage{graphicx}
\usepackage{cite}
\def\BibTeX{{\rm B\kern-.05em{\sc i\kern-.025em b}\kern-.08em
    T\kern-.1667em\lower.7ex\hbox{E}\kern-.125emX}}
\usepackage{balance}
\begin{document}

\title{Mapping network structures and dynamics of decentralised cryptocurrencies: The evolution of Bitcoin (2009-2023)}
\author{Marco Venturini, Daniel García-Costa, Elena Álvarez-García, Francisco
Grimaldo, Flaminio Squazzoni}
\markboth{Formatted to be submitted to Transactions on SMC: Systems}{Mapping network structures and dynamics of decentralised cryptocurrencies: The evolution of Bitcoin (2009-2023)}

\maketitle

\begin{abstract}
Cryptocurrencies have recently been in the spotlight of public debate due to their embrace by the new US President, with crypto fans expecting a 'bull run'. The global cryptocurrency market capitalisation is more than \$3.50 trillion, with 1 Bitcoin exchanging for more than \$97,000 at the end of November 2024. Monitoring the evolution of these systems is key to understanding whether the popular perception of cryptocurrencies as a new, sustainable economic infrastructure is well-founded. In this paper, we have reconstructed the network structures and dynamics of Bitcoin from its launch in January 2009 to December 2023 and identified its key evolutionary phases. Our results show that network centralisation and wealth concentration increased from the very early years, following a richer-get-richer mechanism. This trend was endogenous to the system, beyond any subsequent institutional or exogenous influence. The evolution of Bitcoin is characterised by three periods, \textit{Exploration}, \textit{Adaptation} and \textit{Maturity}, with substantial coherent network patterns. Our findings suggest that Bitcoin is a highly centralised structure, with high levels of wealth inequality and internally crystallised power dynamics, which may have negative implications for its long-term sustainability.
\end{abstract}


\section{Introduction}\label{section1}
Introduced by Satoshi Nakamoto in 2008 \cite{nakamoto2008bitcoin}, Bitcoin is now worth \$1.5 trillion with a daily exchange volume of \$67 billion \cite{Coinmarketcap}, excluding derivatives. In total, the entire cryptocurrency world has a market capitalisation of around \$3.50 trillion. The world's largest asset management firm, BlackRock, launched a spot Exchange Traded Fund (ETF) replicating Bitcoin in January 2024, soon followed by other institutional players. By 6 November 2024, this ETF had a daily exchange volume of \$97 billion \cite{iShares}. The number of crypto users worldwide is estimated to be around 620 million in 2024, rising to 800 million in 2025 \cite{Statista}, Bitcoin transactions per day are at their highest level ever, and 1 Bitcoin is exchanged for more than \$97,000 by the end of November 2024. 
While Bitcoin is increasingly present in the portfolios of retail investors, it has also recently been at the centre of the electoral campaign for the 2024 US presidential election, with the newly elected president promising to make the United States the ``crypto capital of the world''. On the day of Trump's victory, the price of Bitcoin jumped from \$69,500 to around \$76,000, thus marking an all-time high price. 

This shows that Bitcoin is no longer a niche product of a group of idealistic ``hippies'', but is part of a wider political agenda and is being included in financial institutions' derivative products and in portfolio investing strategies. However, Bitcoin is a multi-faceted, complex system where different entities and instances coexist, from exchanges and financial institutions to retail and fraudulent users. For instance, Bitcoin has been instrumental in the creation of SilkRoad, a dark market founded in 2011 and shut down by the Federal Bureau of Investigation (FBI) in October 2013, but also of regulated exchanges such as Coinbase, a secure online platform for buying, selling, transferring and storing cryptocurrencies, now listed on the NASDAQ with a market capitalisation of \$77 billion.

Beyond these bright and dark sides, the secret of Bitcoin's success is that it allows anonymous peer-to-peer transactions without any intermediary, relying solely on its decentralised consensus protocol \cite{mingxiao2017review}. The Bitcoin transaction network consists of several elements: addresses, transactions and blocks. The recipient of a transaction is an address identified by its public key, a univoque alphanumeric sequence. An address can only send what it has previously received: this is known as the UTXO, Unspent Transaction Output, which characterises Bitcoin. Transactions, in turn, are defined as the transfer of tokens from one address to another (\textit{peer-to-peer}); once confirmed, they are broadcast to the network \cite{park_2019} and collected into new blocks by miners who compete to solve a cryptographic puzzle \cite{back2002hashcash} to add the candidate block to the blockchain. This process, known as ``proof-of-work,`` selects miners based on their computational effort \cite{Maesa_al_2018}.

Understanding the complexity of this new type of economic exchange infrastructure requires a detailed study of a complex transaction network, made up of different components that interact in non-trivial way. This challenge has attracted many computational scientists, with studies of the transaction network of Bitcoin since its early years \cite{reid2013analysis, kondor2014rich}. Unfortunately, the research has been hampered by a limited timescale of observation. For instance, \cite{Maesa_al_2018} analysed the evolution of the Bitcoin transaction network and tested the rich-get-richer mechanism using data until December 2015. Similarly, in \cite{The_Anti-Social}, the authors built a longitudinal dataset of the Bitcoin network to study its structural changes over time, but relied on data only up to September 2014. A notable exception is \cite{nerurkar2021dissecting}, in which the authors used data up to May 2020 to provide a comprehensive analysis of Bitcoin network dynamics from January 2009 to the early stages of the Covid-19 pandemic. Furthermore, these studies have tended to focus on specific Bitcoin aspects to capture its network properties, such as looking at degree distributions, power laws or preferential attachment, rather than attempting a comprehensive analysis of its network formation, dynamics and evolution. 
Finally, while previous research has considered a longitudinal perspective to capture network dynamics, it has ignored the impact of certain important exogenous factors, such as law enforcement and public interest, with the exception of \cite{tasca2018evolution}, which linked the prevalence of business entities in the network to the Bitcoin external evolution and perception of Bitcoin, from a prototype to a ``sin'' phase towards legitimate business.

Our study aims to contribute to the computational literature on Bitcoin, and cryptocurrencies in general, by extending the dataset timeframe, using multiple measures, and investigating the network dynamics by identifying different evolutionary stages. First, we used a dataset updated to the latest available year (2023) and adopted a systemic approach, i.e. a large number of measures, to fully characterise the Bitcoin transaction network. We then mapped Bitcoin's network properties over 14 years, focusing in particular on centralisation and concentration trends, significantly extending all previous research time windows. We took a longitudinal approach with annual snapshots of the network and examined the system dynamics. We also tested for the rich-get-richer phenomenon, allowing us to assess whether the centralisation and concentration trends are due to the increasing presence of institutional investors or to the endogenous growth dynamics of Bitcoin since its early stages. Furthermore, we confirmed previous findings on disassortative and clustering tendencies, while also investigating the network connected components and their composition.
Finally, we identified three periods in the evolution of Bitcoin: the \textit{Exploration}, \textit{Adaptation} and \textit{Maturity} phases.

Our study had three objectives. First, we analysed the entire transaction network of Bitcoin. Second, we studied the dynamics of centralisation and concentration by considering the multiple tensions between the presence of institutional actors and the endogenous force of wealth accumulation typical of financial markets. Third, by combining our data with Bitcoin-related events, we reconstructed the evolution of the system by identifying its different phases. Our results can improve our understanding of cryptocurrencies, one of the most important financial innovations of the last decades, and we hope that our methods and results can inspire future computational research.
 
The remainder of the paper is structured as follows. In \hyperref[section2]{Section 2}, we review the existing research on the Bitcoin transaction network and related crypto assets. In \hyperref[methods]{Section 3}, we present our data and methods. \hyperref[results]{Section 4} shows the results of our analysis, while \hyperref[discussion]{Section 5} presents our main findings and discusses the results and limitations of our study.

\section{Background}\label{section2}

Although research on Bitcoin has focused on multiple layers (six according to \cite{taxonomy_centr}), here we focused our review on studies of the structures and dynamics of the Bitcoin transaction network that have considered transactions between addresses. We also discussed the application of this approach to the study of other blockchain-based digital assets and systems, in particular Ethereum and Non-Fungible Tokens (NFTs), to show how these approaches can be extended beyond Bitcoin to other blockchain infrastructures.

In general, research attention has mostly been devoted to fitting power law distributions, using cryptocurrencies as an illustration of complex network dynamics \cite{ barabasi1999emergence, barabasi2009scale}.As the literature has acknowledged the scale-free nature of the Bitcoin network, despite some conflicting results, we used this point as a \textit{fil rouge} to reconstruct the existing research on the topic. Finally, as anticipated in \hyperref[section1]{Section 1}, we focused on the phases of the Bitcoin's life, prioritising research that had already performed similar analyses.

A first study on the transaction network of Bitcoin and its anonymity was proposed by \cite{reid2013analysis}, where they analysed blockchain data up to the 12 July 2011. The authors specified two network configurations, one where transactions were nodes and edges were the flows of tokens, and another one where nodes were users -a contraction of all users appearing together as transaction inputs- and edges were the connection between each input-output of the transaction. In both cases, they found skewed degree distributions, but rejected the hypothesis of a power law fit. They also showed the risks to privacy associated with Bitcoin data management and questioned the effectiveness of anonymity in the system, being able to retrieve the IPs of some of the addresses and trace each of the transactions derived from a Bitcoin theft that occurred on 3 June 2011.

\cite{ron2013quantitative} extended the analysis of the Bitcoin transaction network by considering data up to 13 May 2012. They again relied on two sets of data, a graph of all Bitcoin addresses and transactions, and a compressed version in which users were reconstructed by addresses that appeared as inputs in the same transactions. In a static network representation, the authors found that the vast majority of addresses moved only small amounts of Bitcoin, while a few hundred of them moved more than 50,000 BTC, with 98\% of nodes having less than 10 BTC as their balance. They also found that 78\% of the existing tokens were dormant, accumulating in ``savings accounts'' with only incoming transactions. Furthermore, among the 364 transactions over 50,000 BTC considered in their analysis, the authors focused on one large transaction that moved 90,000 BTC on 8 November 2010 and its descending transfers: they concluded that most of the network's largest transactions were descended from this one.

In a study of the Bitcoin transaction network between 3 March 2009 and 4 October 2013, \cite{baumann2014exploring} found right-skewed distributions that converge to a scale-free network over time. They also found a strong correlation between user activity (and number of transactions), and the USD/BTC exchange rate: each spike in the exchange rate was shortly followed by a spike in activity and transactions. They also found that higher user activity reduced the cliquishness of the network, with dynamics following a small-world-like structure.

\cite{The_Anti-Social} made an important contribution by analysing the Bitcoin network of transactions that took place between 3 January 2009 and 6 September 2014. They created a longitudinal dataset by dividing the observation period into 11 snapshots of six months each. Their results showed that the network was disassortative, almost fully connected with 99,9\% of nodes in the largest connected component, and subject to a densification power law. This means that the expansion of the network was driven by the increase in the number of transactions. They also found that the degree of hub-dominance was inversely proportional to the size of the reference community and that there were high inequalities in terms of wealth in the network.
Note that these findings were also confirmed by research on Ethereum \cite{buterin2014next}. This is the case of \cite{Chen_al_2018}, which studied three different networks: the money flow graph (MFG), the smart contract creation graph (CCG), and the smart contract invocation graph (CIG) from the 30 July 2015, to 10 June 2017. For all networks, the authors found skewed degree distributions that fit a power law, low disassortative and clustering coefficients, and a dominant highly connected component that includes the majority of nodes.

Considering data up to the 23 December 2015, creating 20 snapshots with different minimum value thresholds, from 0.000001 BTC to 100,000 BTC, and using a new algorithm to cluster addresses into users (where all inputs appearing in the same transactions refer to the same user), \cite{Maesa2016, Maesa_al_2018} found that the degree distributions fit a power law with coefficients ranging between 2 and 2.5. They also found clustering tendencies and network densification, as well as high Gini coefficients, confirming the presence of high inequality in the network. In addition, they tested for the rich-get-richer mechanism and found that the richest in terms of balance and in-degree increased their resources and maintained their privileged position over time. Beyond these two papers, the authors also studied the ``bow-tie'' structure of the Bitcoin network \cite{maesa2019bow, maesa2019graph} and characterised the roles of each component by means of their composition in terms of network actors.

More recently, \cite{nerurkar2021dissecting} analysed the Bitcoin transaction network with data up to 8 May 2020 and a longitudinal analysis of annual snapshots.
Their results showed that the degree distributions were mainly scale-free, with a high degree of centralisation and concentration. They also found positive clustering tendencies, uncertain assortative mixing and high centralisation. In addition, they found a large connected component consisting mainly of exchanges, mixing services and mining pools. 

Note that these empirical patterns seem to characterise may other cryptocurrencies, not just Bitcoin. For instance, \cite{motamed2019quantitative} compared the transaction networks of Bitcoin, Ethereum, Litecoin, Dash, and Z-Cash over ten years, from 2009 to 2019. 
Results showed a general disassortative tendency, degree distributions that fit a power law, low densities, and positive clustering. Similarly, consistent patterns emerged in \cite{somin2020network}, which analysed 11,900 trading assets implemented in the Ethereum blockchain and their network patterns from February 2016 to January 2019. The results showed the degree distributions following a truncated power law and converging dynamics.

By relating both preferential attachment and different phases of the Bitcoin growth, \cite{kondor2014rich} studied the structures of the Bitcoin transaction network, as well as the evolution and growth of the system. They considered data from 3 January 2009 to 7 May 2013 and used an accumulated network setting to assess changes in the network structures over time. The authors found right-skewed degree distributions that fit a power law, as well as high Gini coefficients that stabilised around 0.5 for both the degree distributions but close to 1 for the balance. They also found a disassortative tendency, a low but significant clustering coefficient and a sub-linear preferential attachment as the mechanism driving network growth. Finally, in terms of network evolution, they divided the evolution of Bitcoin into two periods: an initial phase (until autumn 2010) and a trading phase after mid-2011. This work inspired \cite{liang2018evolutionary}, who studied the Bitcoin transaction network based on monthly snapshots, adding Ethereum and Namecoin to the analysis. Overall, these results confirmed \cite{kondor2014rich}'s findings, although they disputed that the degree distributions actually followed a power law.

In addition to the two phases identified by \cite{kondor2014rich}, \cite{tasca2018evolution} conducted a study on the dominance of different business categories along the evolution of Bitcoin. The authors studied Bitcoin transactions up to May 2015 clustering individual Bitcoin addresses into 2850 super-clusters representing business entities. They then identified four primary business categories: miners, gambling services, black markets and exchanges. By examining the transaction activity of the clusters, they highlighted distinct patterns of behaviour within these categories: for instance, while transactions between traders and exchanges occurred on average every 11 days and amounted to around 20 BTC, movements involving gamblers were only worth around 0.5 BTC and occurred intra-day. Furthermore, considering the dominance of each category over time, they observed three distinct regimes in Bitcoin: an early prototype phase until 2012, a second phase dominated by “sin” entities such as dark markets until 2013, and a third period marked by a shift from “sin” to legitimate businesses. Finally, they suggested a trend that led to Bitcoin eventually being perceived as a legitimate economic infrastructure.

In summary, we reviewed previous network research on the cryptocurrency markets with a particular focus on Bitcoin. In general, all the networks studied showed a prevalence of power law distributions of degree activity and wealth, coupled with significant levels of inequality. This shows that the Bitcoin (and Ethereum) network appears to be unequal, centralised and close to a scale-free network. Moreover, a recurring feature is the weak disassortative tendency that characterises the transaction networks in the cryptocurrency markets. 

The most obvious limitation of previous research is the lack of recent data to confirm whether the patterns previously found are persistent. Most previous studies have only covered the early years of Bitcoin, rarely reaching 2015. Missing nearly a decade of network expansion and shocks, including, for example, the boom of 2021 and the so-called 'crypto-winter' of 2022, when around \$2 trillion of crypto assets were wiped out, is a serious limitation when trying to understand a rapidly evolving infrastructure such as Bitcoin. with the exception of \cite{nerurkar2021dissecting}, which considered data up to May 2020, there are, to the best of our knowledge, no comprehensive works that can serve as a standard reference in the literature. Furthermore, previous research has considered only a limited number of network measures that must be jointly considered to support a comprehensive analysis of the Bitcoin network.

To fill this gap, we built a dataset from 3 January 2009 to 31 December 2023. This timeframe is three years longer than \cite{nerurkar2021dissecting}'s study and about eight to ten years more than longer than any other study. This allowed us to consider the impact of venture capitalists and corporations on bitcoin transaction patterns, whose presence increased significantly from 2015 onwards. By adopting a longitudinal setting with 15 yearly snapshots of the network, we aligned our empirical strategy with the existing literature \cite{The_Anti-Social, Maesa_al_2018, tasca2018evolution}, and extended the analysis beyond the last notable update study by \cite{nerurkar2021dissecting}. Furthermore, when considering the attributes of edges in previous studies, we noted a relevant point that requires careful consideration: transaction activity and transaction value are usually mutually exclusive, with transaction value often playing a minor role. We believe that disentangling these two aspects is necessary to characterise the flow of Bitcoin, and thus adopted a dual weighting to account for both transaction activity and the absolute value of tokens moved per edge ($w_1$ and $w_2$ in \hyperref[formula3]{Equation 3}).

Furthermore, in line with \cite{Maesa2016} and \cite{Maesa_al_2018}, who tested for the rich-get-richer mechanism until 2015, we tested the same mechanism with new data. This allowed us to update their analysis and examine the dynamics and trajectories of wealth accumulation in a period when cryptocurrencies experienced relevant changes, both in terms of structural stability and external shocks, with greater institutional participation, a global pandemic (i.e., Covid-19), and political regulation.  

Finally, while \cite{kondor2014rich} and \cite{tasca2018evolution} have suggested to distinguish the evolution of Bitcoin into two and three periods, respectively, we wanted to explore new temporal categorisations that better reflect the network measures and the Bitcoin history. Here, we propose a categorisation of three periods -\textit{Exploration, Adaptation} and \textit{Maturity}- as they better reflect the trends observed in our analysis and the Bitcoin history, including important exogenous events. Given that previous research relevant to our purposes has either been limited by the short time span considered \cite{kondor2014rich} or has linked the time periods to idiosyncratic factors, such as the prevalence of each general business on the network over time rather than to patterns of network characteristics \cite{tasca2018evolution}, we believe that our study provides a refined map of Bitcoin's evolution.

\section{Methods and Data}\label{methods}
We retrieved all Bitcoin blockchain data from the 1 January 2009 to the 31 December 2023, where each record collected is a block of transactions with input and output addresses, amounts sent and received, a block number and a timestamp. We also included a dummy indicating the existence of a special transaction -with no input address- called Coinbase, which represents the reward for mining the block. This allowed us to collect all time-ordered exchange events, sending or receiving, that have occurred since the inception of Bitcoin across the entire population of addresses. We used this data to perform a longitudinal network analysis of the complete on-chain transactions. 

Following \cite{kondor2014rich} and \cite{The_Anti-Social}, we reconstructed the address-to-address edges  (the \textit{Address network}, as defined by \cite{WU_al_2021}) and the proportional weights based on the amount of tokens moved by each address, as follows:

\begin{equation}\label{formula1}
v^{(i \to j, n)} = \left( v_{\text{in}}^{(i, n)} - \frac{t_{\text{fee}}^{(n)} \cdot v_{\text{in}}^{(i, n)}}{t_{\text{in}}^{(n)}} \right) \cdot \frac{v_{\text{out}}^{(j, n)}}{t_{\text{out}}^{(n)}}
\end{equation}

where $v^{(i \to j, n)}$, the value, is the weight of the edge -the number of tokens transacted- between nodes $i$ and $j$ in transaction $n$; $v_{\text{in}}^{(i, n)}$ is the amount sent by node $i$ within transaction $n$; $t_{\text{in}}^{(n)}$ is the total volume sent in a transaction $n$ by all the participating addresses, and $t_{\text{fee}}^{(n)}$ is the total fee of transaction $n$. We used the same notation $v_{\text{out}}^{(j, n)}$ and $t_{\text{out}}^{(n)}$ referring to the output, where $j$ refers to the receiving node. 

To obtain a more manageable dataset, we encrypted  input/output addresses and transaction hashes, converting public key addresses to integer IDs to reduce storage requirements and improve usability. We maintained a dictionary mapping the new IDs to the original public keys for traceability.

This allowed us to obtain a dataset of approximately one billion of unique transactions and 11.5 billion directed address-to-address transactions, including all the network activity since the first ever transaction, the Genesis transaction initiated by Satoshi Nakamoto the 3rd January 2009. We obtained an edge list representing directed transactions, with associated timestamps and token amounts (see \hyperref[table1]{Table 1}).

\begin{table}[!t]
\caption{The dataset structure.}\label{table1}
\centering
\resizebox{\ifdim\width>\linewidth\linewidth\else\width\fi}{!}{
\fontsize{14}{14}\selectfont
\begin{tabular}[t]{ccccccc}
\hline 
block\_number & transaction\_id & is\_coinbase & input\_address\_id & output\_address\_id & value & timestamp\\
\hline \hline
68726 & 99634 & 0 & 439060846 & 243129826 & 1.7e+07 & 2010-07-17 15:58:16 UTC\\
70737 & 43914 & 0 & 123937997 & 195528996 & 5.0e+06 & 2010-07-28 04:14:14 UTC\\
92069 & 74766 & 0 & 600769453 & 930678510 & 2.1e+07 & 2010-11-15 23:22:22 UTC\\
92104 & 216312 & 0 & 823884578 & 1032404731 & 1.0e+06 & 2010-11-16 04:08:25 UTC\\
\hline
\end{tabular}}

\end{table}

We conceived the Bitcoin system as a directed network 
\( \textit{G} = \left\{\textit{V}, \textit{E}\right\}\) where \textit{V} and \textit{E} are the sets of nodes and edges, respectively. Following \cite{nerurkar2021dissecting}, we considered ``value'' and ``timestamp'' as edge attributes. Since multiple transactions can occur between the same addresses, $G$ is a directed multi-graph. Furthermore, since $G$ is a multi-graph, we were able to group all the transactions going in the same direction under the same pair of nodes. It follows that ``value'', the attributes of the edges indicating the amount of tokens transferred, becomes the arithmetic sum of all the edges within a pair of nodes per year. To account for the otherwise missing in- and out-movements, we developed a second edge attribute called ``activity'', which counts the number of times each transaction between the same pair of nodes is repeated in a year. Our graph object is formulated as follows:

\begin{equation}\label{formula2}
  \textit{G}_t = \left\{\textit{V}_t, \textit{E}_t, \textit{w}_1: E_t \to \mathbb{R}_{> 0}, \textit{w}_2: E_t \to \mathbb{N^*} \right\}  
\end{equation}

where $G_t$ is a weighted graph with only positive weights $w_1$ and $w_2$, corresponding to ``value'' and ``activity'', respectively. In our empirical setting, we choose fifteen consecutive snapshots of the network, \( \textit{t} = \left\{1, \ldots, n \right\}, \textit{n}=15 \), each corresponding to a calendar year, thus capturing the dynamics and evolving patterns that would otherwise be flattened into a static representation.

Given the variation in transaction size and the presence of noise, we set a threshold below which transactions were excluded from the analysis. We set it at $0.0001$ bitcoins per year, meaning that any edge that moves less than (or equal to) $0.0001$ bitcoins between two addresses in a full year is removed from the network. This allowed us to solve two problems: first, we removed so-called ``dust'' transactions \cite{loporchio2023bitcoin}, which are either unintentional or deliberate attacks on an address; second, we did not consider annual movements that have no economic relevance. We performed robustness checks to confirm our choice and the results were consistent (for details, see the \hyperref[results]{Results Section}). The network can be finally defined as follows:

\begin{equation}\label{formula3}
\begin{split}
    &H_t = \{V'_t, E'_t, w_1: E'_t \to \{ w_1 \in \mathbb{R} \mid w_1 > 0.0001 \}, \\
    &w_2: E'_t \to \mathbb{N^*}\}
\end{split}
\end{equation}

where $H_t$ is a subset of the weighted graph $G_t$, with weight $w_1$ only greater than 0.0001 bitcoins. Respectively, $V'_t$ and $E'_t$ are the subset of nodes and edges at time $t$.

We have also removed Coinbase transactions and self-loops generated by the address change operation from the network for the sake of consistency. These are system artifacts that have no substantive meaning when the focus is on the flow of tokens between different addresses.

As we wanted to provide a comprehensive overview of the system evolution, we measured both global and local network metrics. These metrics include degree distributions, connected components, assortative mixing tendencies and Gini indices to assess inequality across distributions. As centralisation and concentration trends are the main focus of our research, we also examined the rich-get-richer pattern \cite{Maesa_al_2018}. Following \cite{Maesa_al_2018, Maesa2016}, we tested two hypotheses. First, the richest nodes increase the share of wealth and in-degree activities they control. Second, those who control most of the wealth and in-degree activities tend to remain the same over time. We adapted \hyperref[formula4]{Equation (4)} and \hyperref[formula5]{Equation (5)} to our empirical case as follows: since $b(v)$ represents the cumulative balance of a node $v$ over time \( \textit{t} = \left\{1, \ldots, n \right\}, \textit{n}=15
\), we calculated the annual balance by taking the total amount received in Bitcoin, then subtracting the total amount sent, adding any Coinbase transactions ($\beta_t(v)$), and subtracting the annual fees paid ($\alpha_t(v)$). The last two terms -Coinbase transactions and annual fees- are implicitly included in our initial edge formulation. Finally, we also added the cumulative wealth from the previous years to the calculation ($\theta_{t-1}(v)$) as follows:

\begin{equation}\label{formula4}
\begin{split}
    &b_t(v) = \sum_{(u, v) \in E'_t} w_1(u, v) - \sum_{(v, u) \in E'_t} w_1(v, u)+ \\ 
    &- \alpha_t(v) + \beta_t(v) + \theta_{t-1}(v)
\end{split}
\end{equation}

Similarly, the in-degree richness $i(v)$ of a node $v$ accounts for the cumulative in-degree at time $t$, where the first term is the in-degree at $t$ and $\gamma_{t-1}(v)$ is the accumulated in-degree of node $v$ until $t-1$ as follows:

\begin{equation}\label{formula5}
i_t(v) = \sum_{(u, v) \in E'_t} w_2(u, v) + \gamma_{t-1}(v)
\end{equation}

Here, unlike the previous part of the analysis, we have included both Coinbase transactions and self-loops, as the focus shifted from the exchange between addresses to wealth accumulation. This was the same test as \cite{Maesa_al_2018}, but with data after 2015.As mentioned above, the system has evolved significantly since then, with the entry of new institutional actors and private investors who have challenged the status quo and accumulated knowledge.

The first hypothesis is tested by considering the ten ($k$) richest nodes, in terms of balance and in-degree, in the set $V$ at time $t$. We used two ratios consisting of the wealth (or in-degree) of the ten richest addresses over the total wealth (or in-degree) of the network. The higher the $r(b)$ and $r(i)$, the higher the inequality with respect to the full set of nodes (see \hyperref[formula6]{Equation (6)}.

\begin{equation}\label{formula6}
\begin{aligned}
    r_t(b) = \frac{\sum_{v \in B_{k,t}} b_t(v) / k}{\sum_{v \in V_t} b_t(v) / |V_t|},
    r_t(i)  = \frac{\sum_{v \in I_{k,t}} i_t(v) / k}{\sum_{v \in V_t} i_t(v) / |V_t|}
\end{aligned}
\end{equation}

The second hypothesis is evaluated by measuring the variability within the set of the ten ($k$) richest nodes, specifically, using union sets. We have calculated the cumulative number of addresses in the set of the richest node at time $t$ over all the years of observation: the theoretical maximum is given by $k * t$ which amounts to $10*15=150$. We have calculated the quantities of interest, $X_{b,t}$ and $Y_{i,t}$ in \hyperref[formula7]{Equation (7)}, as in \cite{Maesa_al_2018}, where $B_{k}^{j}$ and $D_{k}^{j}$ are, respectively, the set of the $k$ richest nodes, at time $t$, in terms of balance and in-degree activity. If both quantities are consistently found below the theoretical maximum, then, our hypothesis is confirmed. Note that we have used the same notation as \cite{Maesa_al_2018} to improve comparability.

\begin{equation}\label{formula7}
\begin{aligned}
    X_{b,t} &= \left| \bigcup_{j=1}^{t} B_{k}^{j} \right|,
    Y_{i,t} &= \left| \bigcup_{j=1}^{t} D_{k}^{j} \right|
\end{aligned}
\end{equation}

\section{Results}\label{results}

This section presents our results with particular emphasis on both global and local measures, and the test of the ``rich-get-richer'' pattern.

\subsection{Global and Local measures}

We started by characterising the network growth over time. \hyperref[table2]{Tables 2} and \hyperref[table3]{3} show the number of nodes and edges for each of the fifteen snapshots considered. The growth of the network is rapid, starting with 2873 nodes and 3500 edges in 2009 and reaching 148 million nodes and 568 million edges in 2023. The scale of this expansion is remarkable and highlights the success that Bitcoin has had over the years. Given the size of the network, its low density is not very surprising: a network with a billion of unique nodes is rarely dense. Interestingly, \hyperref[table2]{Tables 2} and \hyperref[table3]{3} show that edge density stabilised after 2015, indicating that the system reached a more stable number of transactions per node. This is confirmed by the mean degree (\hyperref[figure3a]{Figure 3a}), which stabilises after the same year, probably due to the increased maturity of Bitcoin. 

\begin{table}[!t]
    \centering
    \caption{Summary statistics of the network snapshots: number of addresses and edges from 2009 to 2016. edge density is also shown for each year.}\label{table2}
    \resizebox{\ifdim\width>\linewidth\linewidth\else\width\fi}{!}{
\fontsize{14}{14}\selectfont

    \begin{tabular}{ccccccccc}
    \hline
        Year & 2009 & 2010 & 2011 & 2012 & 2013 & 2014 & 2015 & 2016 \\ \hline \hline
        \# Address & 2873 & 122183 & 2320838 & 5948985 & 15990181 & 34185739 & 55780129 & 94962579 \\ 
        \# Edges & 3500 & 176946 & 5477578 & 16944653 & 52174726 & 240273864 & 297597762 & 314446217 \\ 
        Density & 4.24e-05 & 1.19e-08 & 1.02e-08 & 4.79e-07 & 2.04e-07 & 2.06e-06 & 9.56e-06 & 3.49e-06 \\ \hline
    \end{tabular}}
\end{table}

\begin{table}[!t]
    \caption{Summary statistics of the network snapshots: number of addresses and edges from 2017 to 2023. edge density is also shown for each year.}\label{table3}
    \centering
    \resizebox{\ifdim\width>\linewidth\linewidth\else\width\fi}{!}{
\fontsize{14}{14}\selectfont
    \begin{tabular}{cccccccc}
    \hline
        Year & 2017 & 2018 & 2019 & 2020 & 2021 & 2022 & 2023 \\ \hline \hline
        \# Address & 143253859 & 117644156 & 130922321 & 166858661 & 167060474 & 151005961 & 148245334 \\ 
        \# Edges & 560461106 & 430264236 & 565379186 & 658268828 & 597585938 & 542948094 & 567921141 \\ 
        Density & 2.73e-06 & 3.11e-06 & 3.30e-06 & 2.36e-06 & 2.14e-06 & 2.38e-06 & 2.58e-06 \\ \hline
    \end{tabular}}
    
\end{table}

In \hyperref[methods]{Section 3}, we filtered the data to avoid noise in the network and to obtain more consistent results. We ran some tests to confirm that filtering out the transactions with a cumulative annual value of less than 0.0001 bitcoins did not bias our analysis. \hyperref[table4]{Table 4} shows the share of the Bitcoin volume considered with the filter to that of the whole network. The values range between 0.99 and 1, meaning that we have (approximately) considered the total Bitcoin volume even when applying the filter. Similarly, the share of nodes considered indicates that we were able to account for the vast majority of nodes in the network after removing noise and non-economically significant transactions. Note that the ``NA'' values are due to the remarkable size of the dataset for these specific years and computational issues, but we are confident that the numbers are consistent with other years.

\begin{table}[!t]
    \caption{Share of the Bitcoin volume and nodes considered when filtering out transactions with an annual cumulative value of less than 0.0001 BTC.}\label{table4}
    \centering
    \resizebox{\ifdim\width>\linewidth\linewidth\else\width\fi}{!}{
\fontsize{20}{20}\selectfont
    \begin{tabular}{cccccccccccccccc}
    \hline
        Year &2009 & 2010 & 2011 & 2012 & 2013 & 2014 &2015 & 2016 & 2017 & 2018& 2019& 2020 & 2021& 2022& 2023 \\ \hline \hline
        BTC Considered & 1 & 1 & 1 & 1 & 0.99 & 0.99 & 0.99 & 0.99 & 0.99 & 0.99 & 0.99 & 0.99 & 0.99 & 0.99 & NA  \\ 
        Nodes Considered & 1 & 0.99 & 0.89 & 0,99 & 0.97 & 0.98 & 0.96 & 0.98 & 0.96 & 0.92 & 0.95 & 0.96 & NA & 0.94 & NA  \\ \hline
    \end{tabular}}
    
\end{table}

\begin{figure}[!t]
\centering
\includegraphics[width=3in]{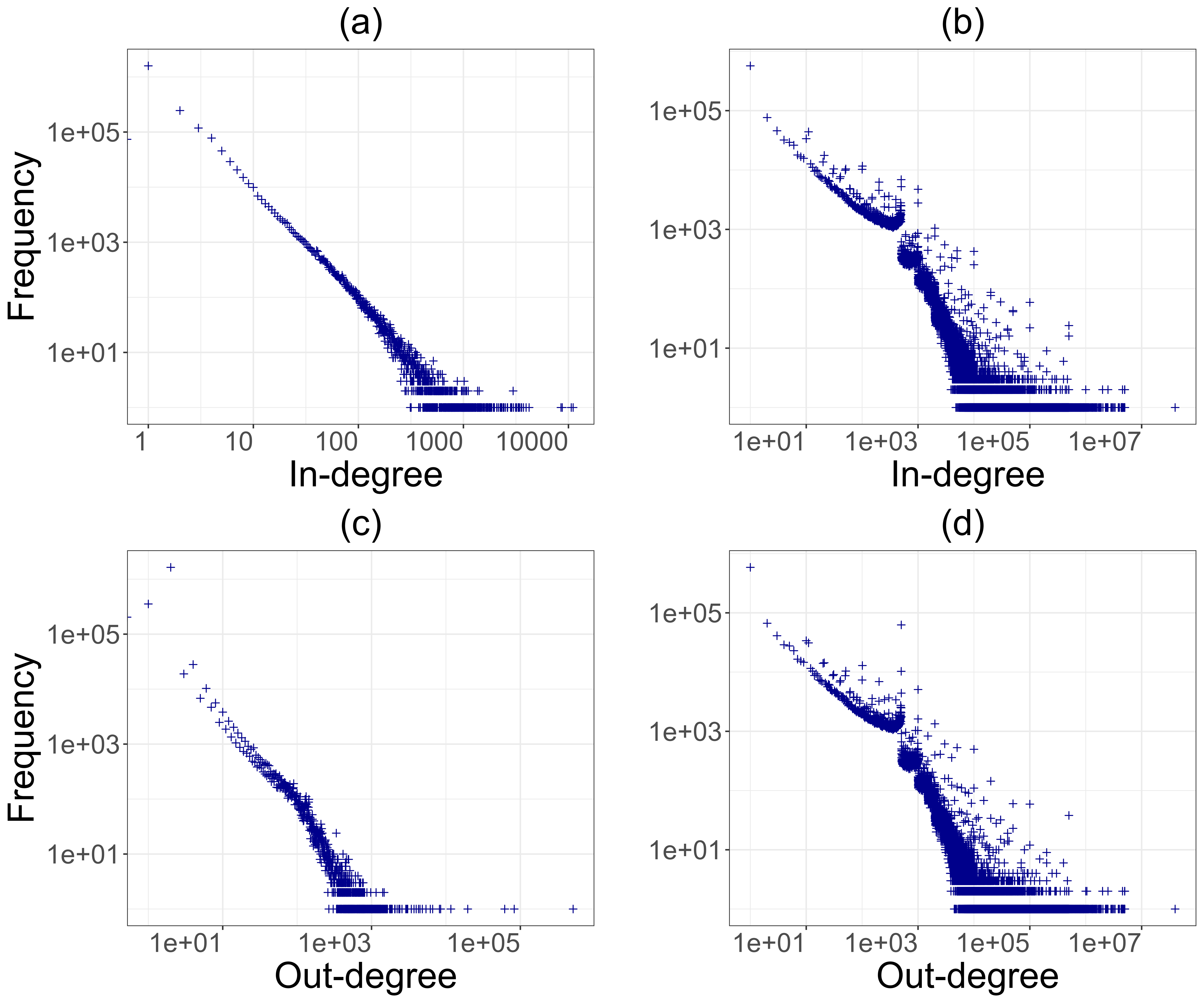}
\caption{The four degree distributions in 2011: (a)in-degree weighted by activity, (b) in-degree weighted by value,(c) out-degree weighted by activity, (d) and out-degree weighted by value.}
\label{figure1}
\end{figure}

\begin{figure}[!t]
\centering
\includegraphics[width=3in]{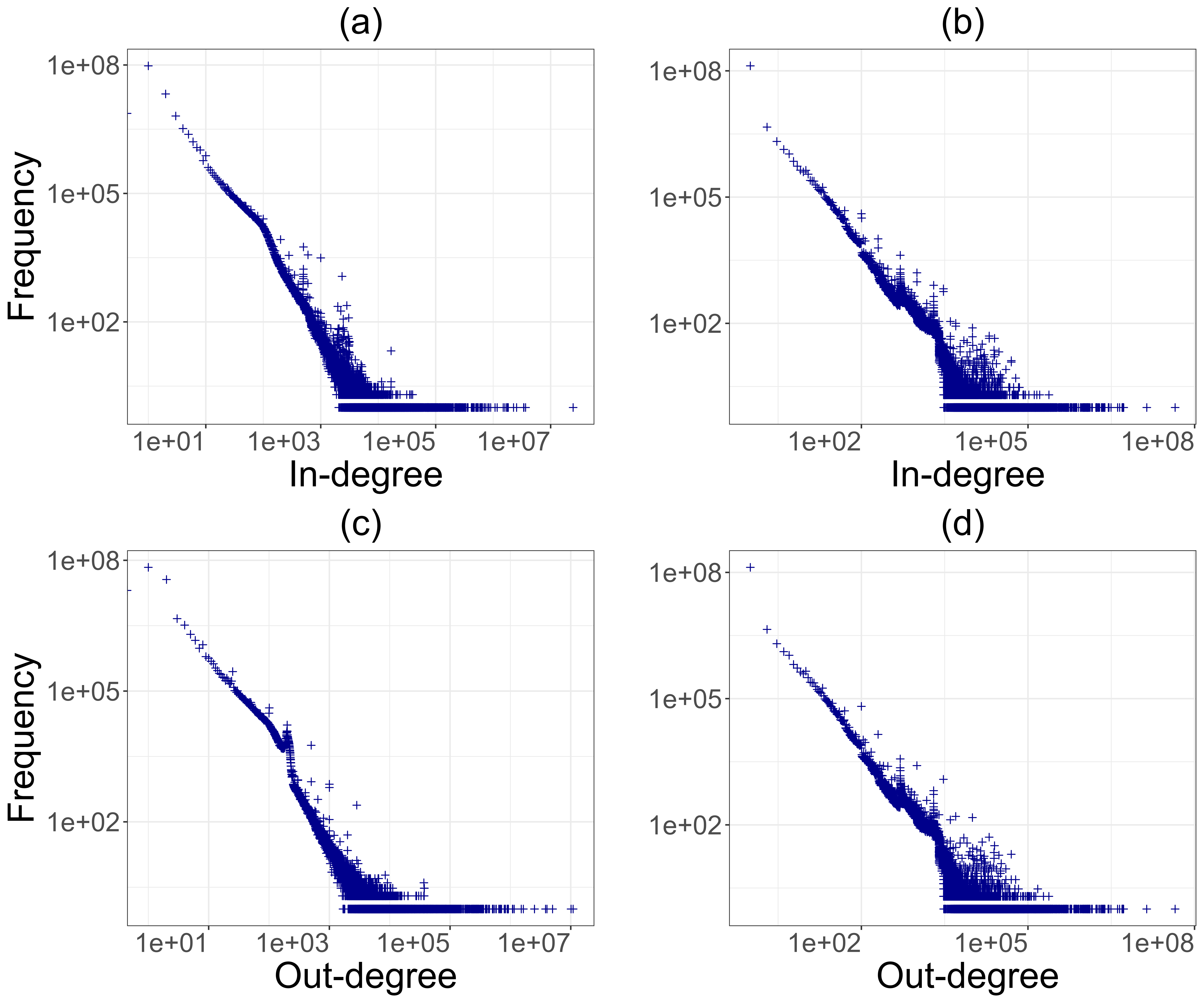}
\caption{The four degree distributions in 2023: (a) in-degree weighted by activity, (b) in-degree weighted by value, (c) out-degree weighted by activity, and  (d) out-degree weighted by value.}
\label{figure2}
\end{figure}

\begin{figure*}[!t]
\centering
\subfloat[]{\includegraphics[width=2.5in]{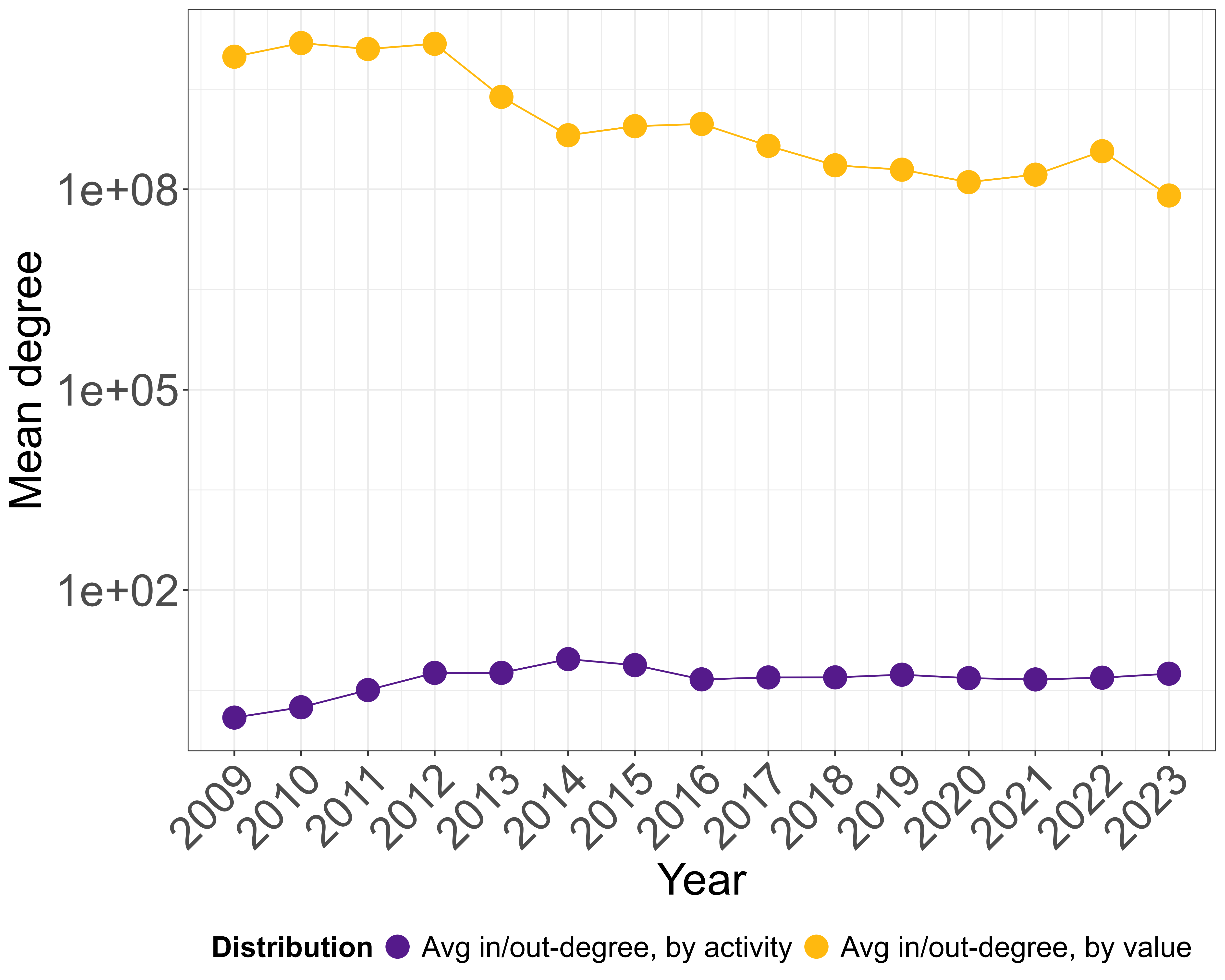}%
\label{figure3a}}
\hfil
\subfloat[]{\includegraphics[width=2.5in]{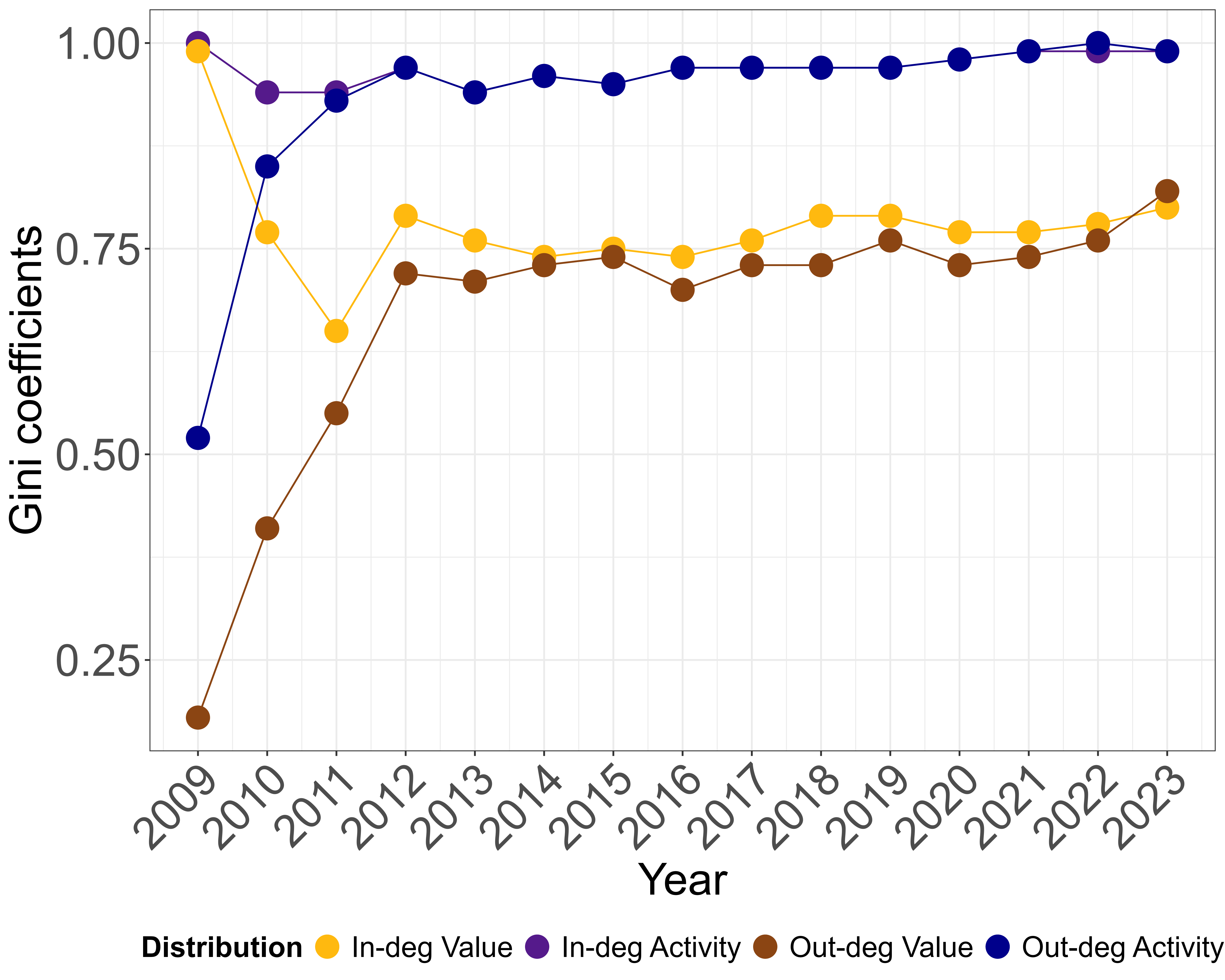}%
\label{figure3b}}
\caption{(3a) Average degree weighted by activity and value. In this case, in- and out-degree averages are equal. (3b) Gini coefficients of the four degree distributions.}
\label{figure3}
\end{figure*}

We also computed the in-degree and out-degree of nodes weighted by activity and value, as specified in \hyperref[methods]{Section 3}, along with the respective distributions. To test for network centralisation and concentration, we examined the degree distributions. \hyperref[figure1]{Figure 1} and \hyperref[figure2]{Figure 2} show two different years, one representing the early years of activity and the other the last period of observation. The four distributions for 2011 and 2023 show a clear pattern that has emerged since the early stages of the system's development: transaction volume is concentrated in the hands of a small number of addresses, while the majority of nodes only manage a few exchanges per year. This trend remains stable throughout the period up to 2023 (\hyperref[figure2]{Figure 2}), despite the growing size of the network. 

We also looked at the higher moments of the distributions: the average activity-weighted degree has stabilised since 2015, while the average value-weighted degree has decreased as the price of Bitcoin has risen, and the standard deviations followed the trends. The decrease occurred because, while Bitcoin is still made up of the same units -satoshi- the unit prices have increased over time, reducing the number of tokens that need to be moved to transfer the same value in dollar terms. Skewness and kurtosis have increased over time, indicating increasing centralisation, unevenness and heavier tails, to stabilise again after 2015. It is important to note that the observed centralisation does not indicate a unique evolutionary process, but rather mimics trends common in other complex network infrastructures \cite{barabasi1999emergence, albert2002statistical}, such as technological networks. 

To measure the inequality in the network, we computed Gini coefficients for the four degree distributions: all were above 0.75 after 2011-2012, with those associated with activity-weighted distributions approaching 1. \hyperref[figure3b]{Figure 3b} shows higher Gini values for the two distributions related to activity, i.e., the number of transactions sent or received. We might have expected the opposite because, while completing a transaction is easy and relatively inexpensive, moving a significant amount of tokens is more difficult and costly. This pattern can be explained by the large number of transactions that large addresses complete each year; they may require software or facilities that smaller participants do not have access to, which explains the greater inequality between the two distributions. Finally, due to the small number of addresses and the popularity of Coinbase transactions, which accounted for the majority of transactions at the time, the out-degree distributions in 2009 have lower Gini coefficients than their in-degree counterparts.

\begin{figure*}[!t]
\centering
\subfloat[]{\includegraphics[width=2.5in]{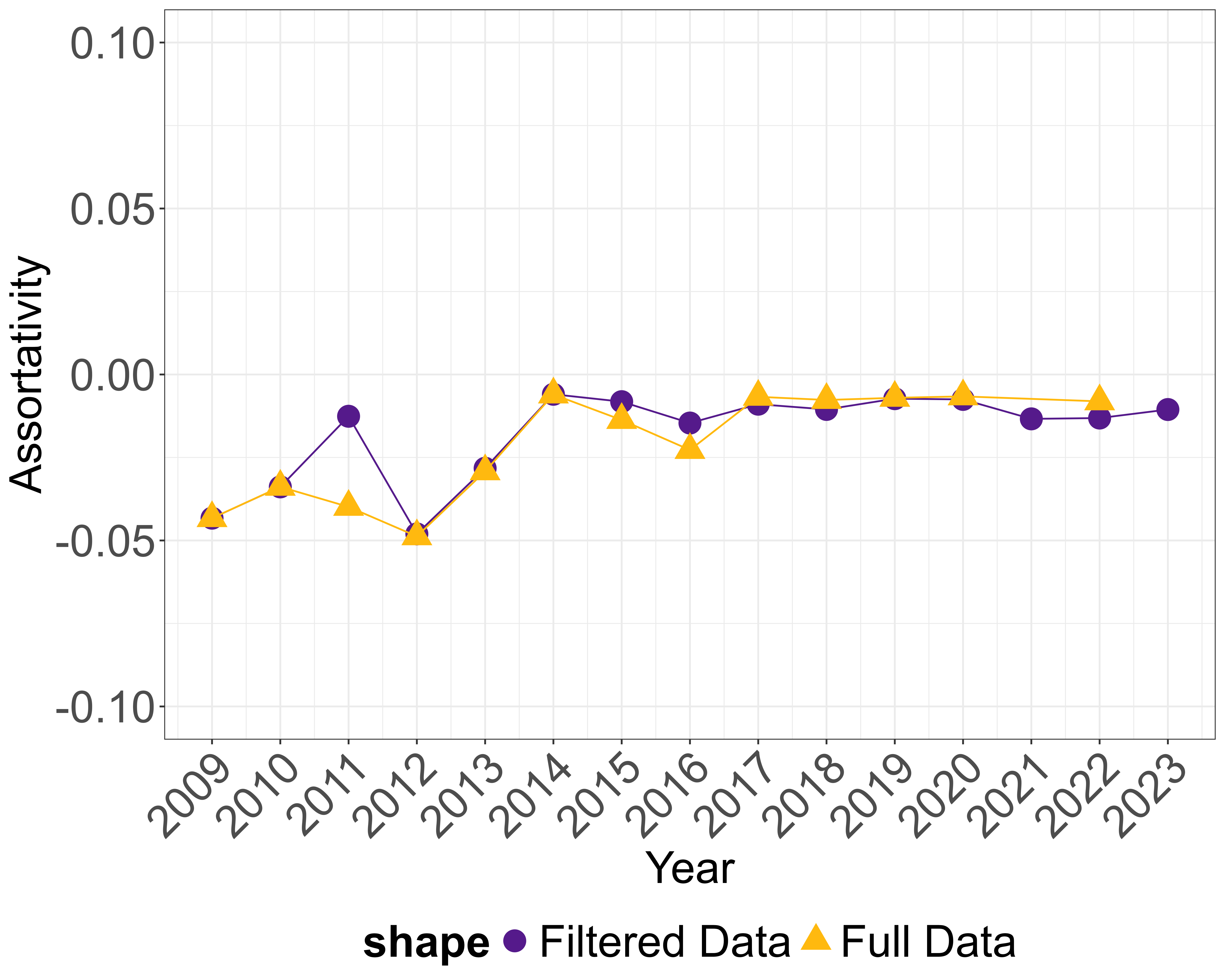}%
\label{figure4a}}
\hfil
\subfloat[]{\includegraphics[width=2.5in]{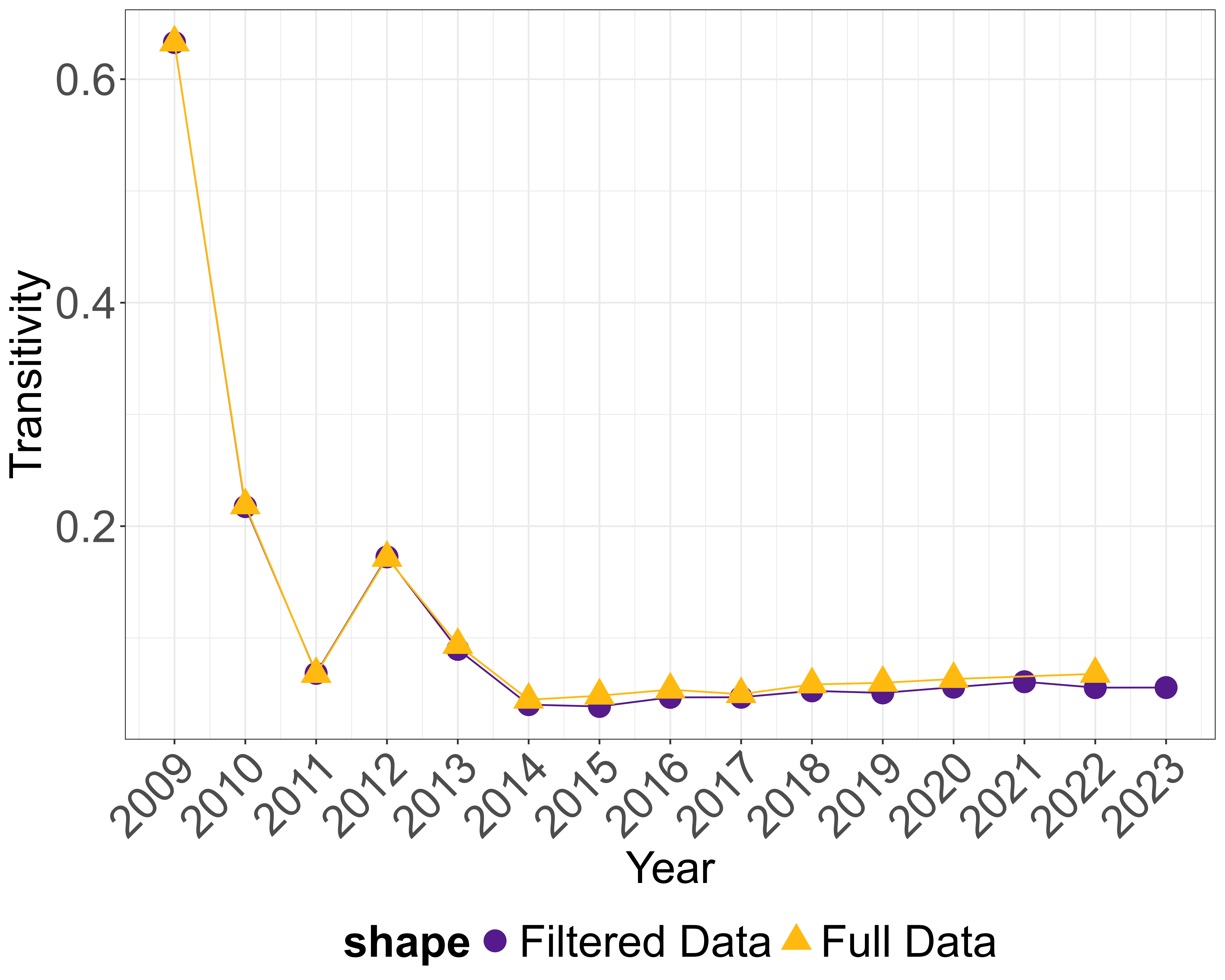}%
\label{figure4b}}
\caption{Effect of the filter on Degree Assortativity and Transitivity Coefficients, per year.}\label{figure4}
\end{figure*}

\hyperref[figure4]{Figure 4a} shows the results of an analysis of assortative mixing \cite{newman2002assortative, newman2003mixing} and clustering motifs to track homophily and clustering within the network. Specifically, we computed degree assortativity and the transitivity index. We found that the degree assortativity coefficient was low but negative over the years, and stabilised again after 2015. This indicates a general tendency for high degree nodes to be connected to nodes that transact with fewer other nodes.

\hyperref[figure4]{Figure 4b} shows the transitivity coefficient, which we measured using the average local clustering coefficient to avoid biased measures due to network sparseness. The coefficient was found to be positive, albeit small, and to stabilise after 2014, consistent with previous findings \cite{kondor2014rich} in both magnitude and direction. Such transitivity suggests two possibilities: either addresses tend to form triadic structures, or the network operates as a closed system in which tokens circulate among the same groups of nodes. 

We also plotted the assortativity and transitivity indices for the full data (as in \hyperref[methods]{Section 3}). \hyperref[figure4]{Figure 4} shows that the filtered and full measures are consistent across the two different datasets. If anything, the small variation we observe assortativity in 2011 strengthens our strategy. This means that we have not overestimated any outcome; at most, we have underestimated some effects in the early years of the network. This confirms the effectiveness of our filter.

We also examined the connected components of the network $G$, distinguishing between ``weakly'' and ``strongly'' connected components, where a weakly connected component disregards edge directions, while a strongly connected component considers them \cite{borgatti2011network}. We found that the largest weakly connected component (LWCC) includes almost all addresses since 2011, indicating that approximately every node can be reached by every other node by either receiving or sending tokens. Thus, despite the sparsity of the network, every node is indirectly connected to the entire set of nodes. 

\begin{figure*}[!t]
\centering
\subfloat[]{\includegraphics[width=2in]{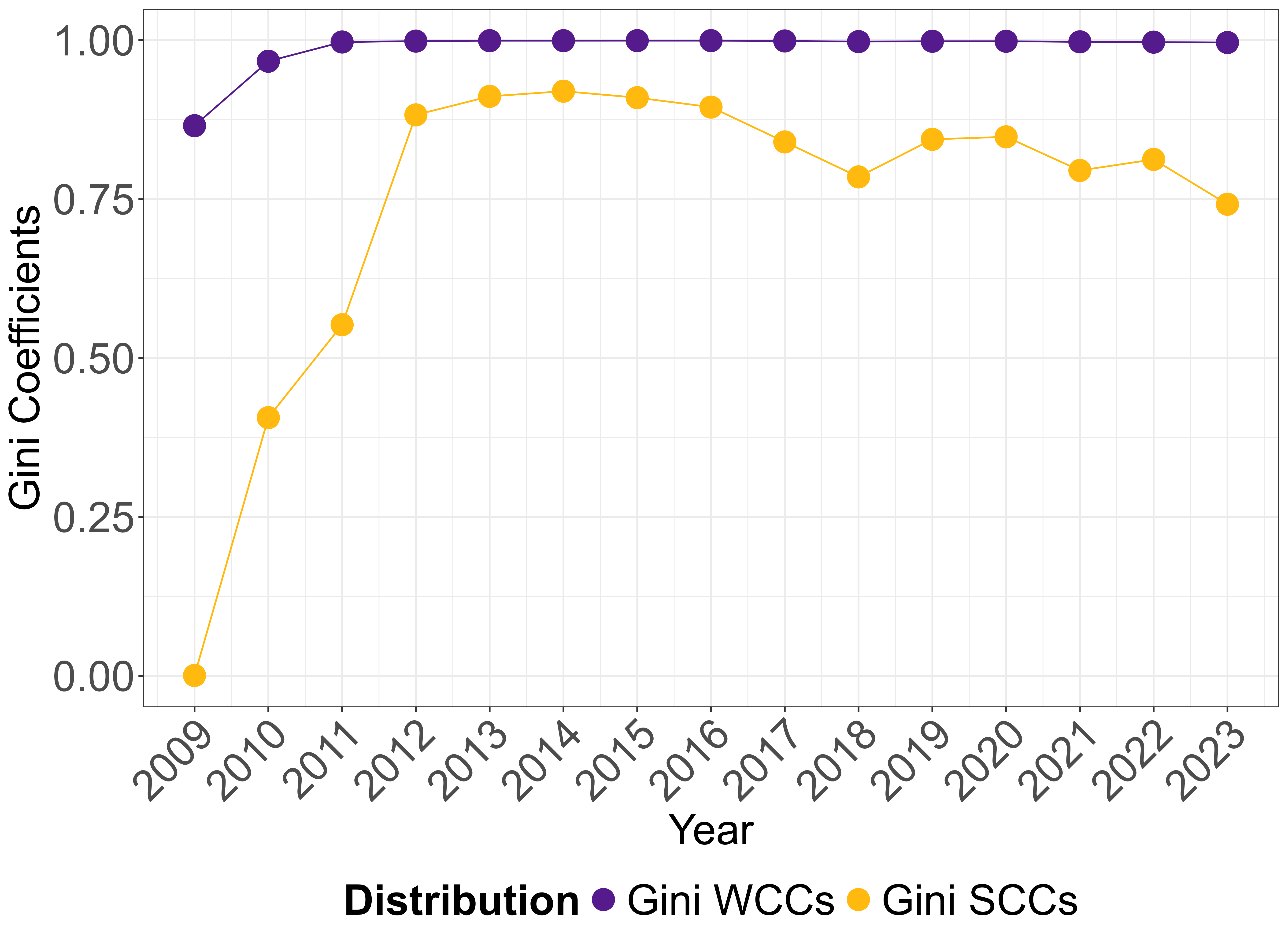}%
\label{figure5a}}
\hfil
\subfloat[]{\includegraphics[width=2in]{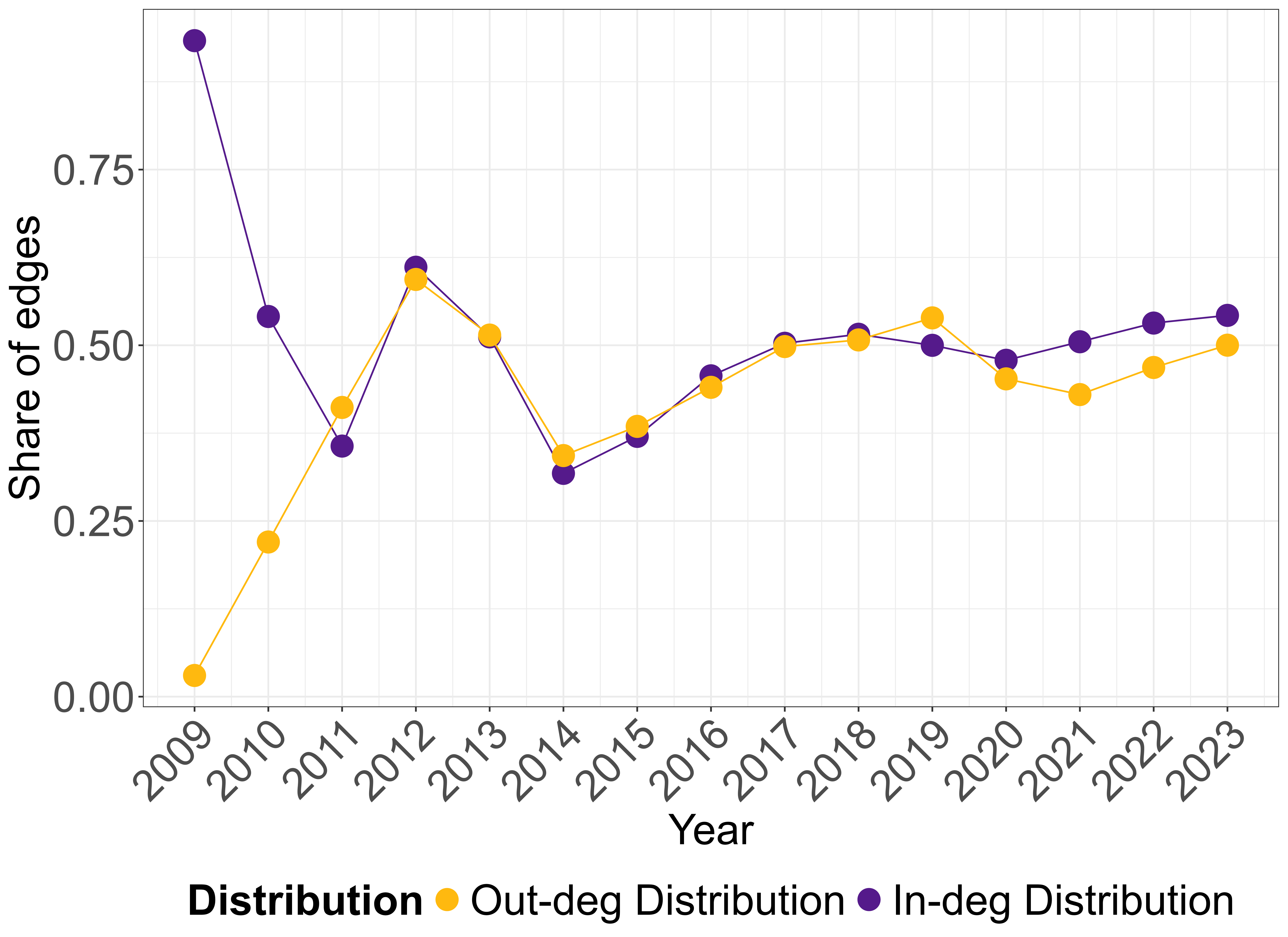}%
\label{figure5b}}
\hfil
\subfloat[]{\includegraphics[width=2in]{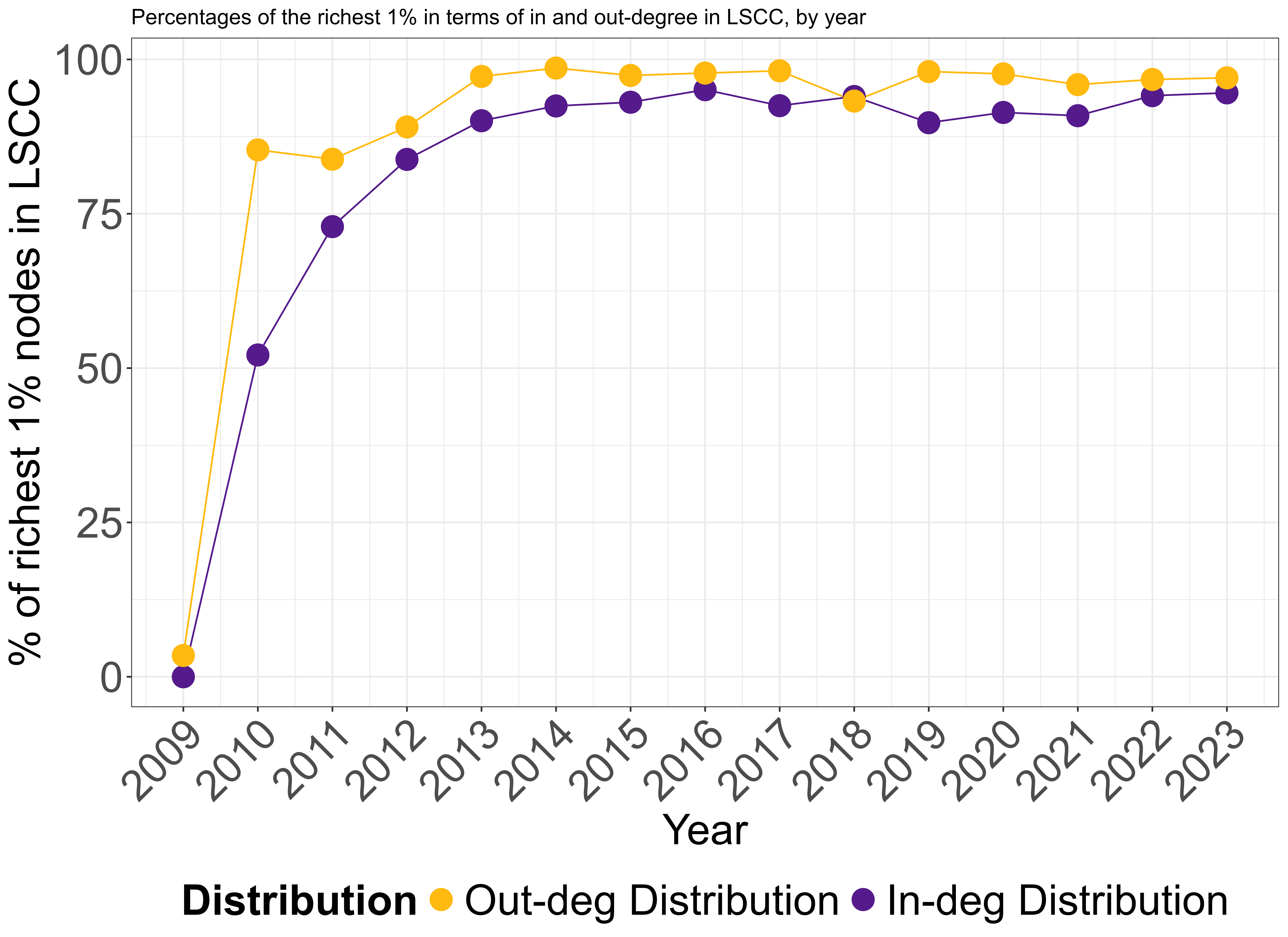}%
\label{figure5c}}
\caption{(a) Gini Indices of the size distribution of the Weakly and Strongly Connected Components. (b) Shares of the in- and out-egdes controlled by the richest 1\% nodes in terms of in- and out-degree. (c) Percentages of the richest 1\% in terms of in and out-degree that are present in the Largest Strongly Connected Component.}\label{figure5}
\end{figure*}

On the other hand, the Largest Strongly Connected Component (LSCC) does not include all addresses but a significant proportion of nodes. The proportion of nodes in the LSCC peaked in 2014, the year of the Mt. Gox crisis, and has since fluctuated inversely with the annual returns for 2015-2017 and 2021-2023, while correspondingly with the returns for 2018-2020. To corroborate our findings on the LSCC and LWCC, we computed Gini indices on the size distributions of the two types of connected components. \hyperref[figure5]{Figure 5a} shows that the Gini index for the WCCs has been consistently close to 1 since the network's inception, indicating a high degree of inequality in the size distribution. Similarly, the Gini index for the SCCs remains above 0.75, albeit with fluctuations that mirror the behaviour of the LSCC.

It is worth noting that the network lacks any relevant completely separate component or multiple centres of activity. Indeed, everything revolves around a giant component that is entirely reachable through indirect connections: a feature that was neither explicitly designed nor anticipated. In addition, there is a substantial strongly connected component that tends to shrink as Bitcoin's value increases.

\hyperref[figure5]{Figure 5b} shows the $1\%$ of the nodes with the highest in-degree and out-degree, including the share of their controlled in-edges and out-edges. Despite fluctuations, these shares converge around 0.5, indicating that the most active $1\%$ control about half of the network's transactions. This concentration peaked between 2011 and 2012, declined until 2014, and increased from 2015 onwards, reducing the range of oscillations. These years were significant: the former marked the end of the exploratory phase, while the latter represents a pivotal moment following the shutdown of SilkRoad by the FBI in October 2013 and the crisis of Mt. Gox, the oldest and once largest cryptocurrency exchange, in 2014, leading to a more mature phase starting in 2015.

We also measured the annual percentage of these $1\%$ of nodes present in the LSCC and LWCC over time(see \hyperref[figure5]{Figure 5c}). For the LSCC, these percentages were initially low, but increased significantly, peaking in 2014 and stabilising thereafter. The percentage of the richest nodes by out-degree in the LSCC was higher than that by in-degree, indicating that the LSCC includes more of the most active out-degree nodes (senders) rather than in-degree nodes (receivers). This is expected, as high out-degree nodes are crucial for network connectivity, while high in-degree nodes may receive most of their information from a few other nodes and remain outside the strongly connected component.

Finally, we identified three main periods in the life of the Bitcoin network: from 2009 to 2012 the system was in the \textit{Exploration} phase, from 2012 to 2015 in the
\textit{Adaptation} phase and from 2015 in the phase of \textit{Stabilisation} phase. This periodisation is based on two different aspects, namely the trends of the measures we have presented and the timing of the main Bitcoin events. On the one hand, most of the measures we observed fluctuated strongly in the Exploration phase, showed some rapid changes in the Adaptation phases and tended to stabilise from 2015 onwards. On the other hand, from the beginning until 2012, the system was in its early phase with few participants and Bitcoin only reached a dollar value in 2011; 2012 marked the end of this phase with the first halving of Bitcoin. Between 2012 and 2015, several events characterised Bitcoin: SilkRoad was shut down by the FBI in 2013, Mt. Gox went bankrupt in February 2014, and in the same year the US Internal Revenue Service  started to consider Bitcoin-related gains and losses as reportable assets; in this phase, Bitcoin adapted to greater participation and external influence. Finally, we identified the maturity phase as beginning in 2015, when volatility, while still significant, gradually decreased, institutional participation increased, and Bitcoin began to shed its stigma as a vehicle for illicit activity. 

\subsection{Rich-get-richer}\label{methods.rich}

For the rich-get-richer mechanism, where we expected the richest nodes to accumulate wealth and activity, we tested \textit{hypothesis 1} and \textit{hypothesis 2} as detailed in \hyperref[methods]{Section 3}. We first examined the increase in wealth controlled by the richest ten ($k$) nodes over time, and then the change in the composition of the set of richest nodes. We found substantial evidence to support both hypotheses: the richest nodes increased their share of wealth and in-degree activity, while the addresses controlling the most money and in-degree activity remained relatively stable.

To calculate $r_t(b)$ and $r_t(i)$ (see \hyperref[formula6]{Equation 6}), the wealth (or in-degree) controlled by the $k$ nodes relative to the whole network, and to measure $X_{b,t}$ and $Y_{i,t}$ (see \hyperref[formula7]{Equation 7}), indicating the variation in the richest set, we de-anonymised most of these addresses using data from BitInfoCharts, Arkham Intelligence, and the datasets from \cite{JBWD_ICDM18, JBWD_CVPR19}.

\begin{figure*}[!t]
\centering
\subfloat[]{\includegraphics[width=2.5in]{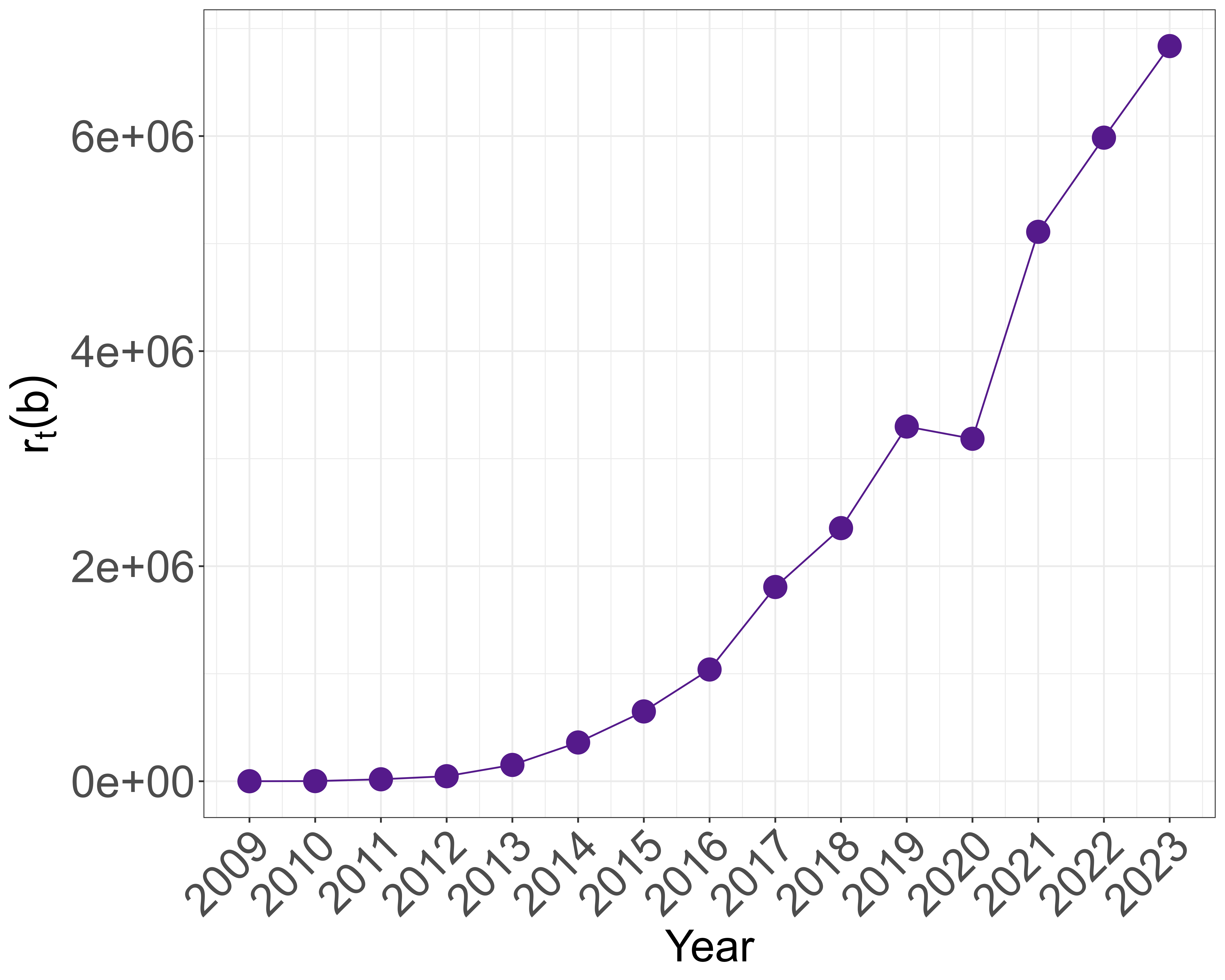}%
\label{figure6a}}
\hfil
\subfloat[]{\includegraphics[width=2.5in]{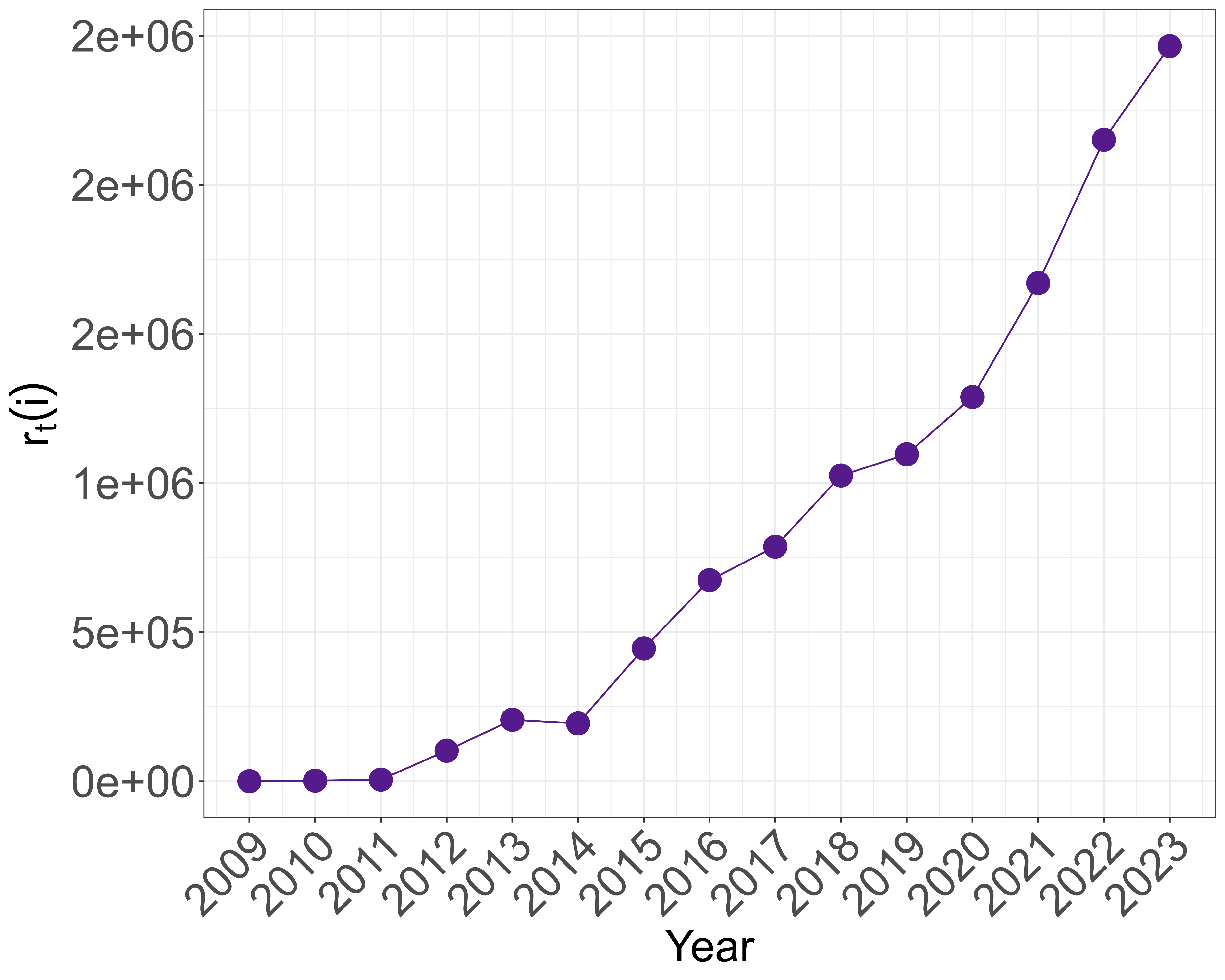}%
\label{figure6b}}
\caption{(a) Evolution of $r_t(b)$ and (b) $r_t(i)$ over time; $r_t(b)$ and $r_t(i)$ are defined as in \hyperref[formula6]{Equation 6}.}\label{figure6}
\end{figure*}

The curves shown in \hyperref[figure6]{Figure 6} reveal a clear pattern of growth in the ratios $r_t(b)$ and $r_t(i)$ over time, thus, indicating an increasing concentration of wealth among the ten richest nodes, both in terms of wealth and in-degree. Specifically, \hyperref[figure6]{Figure 6}, shows that the set of $k$ nodes is richer at time $t$ than at $t-1$ with respect to the whole network. Therefore, \textit{Hypothesis 1} is confirmed.

\begin{figure*}[!t]
\centering
\subfloat[]{\includegraphics[width=2.5in]{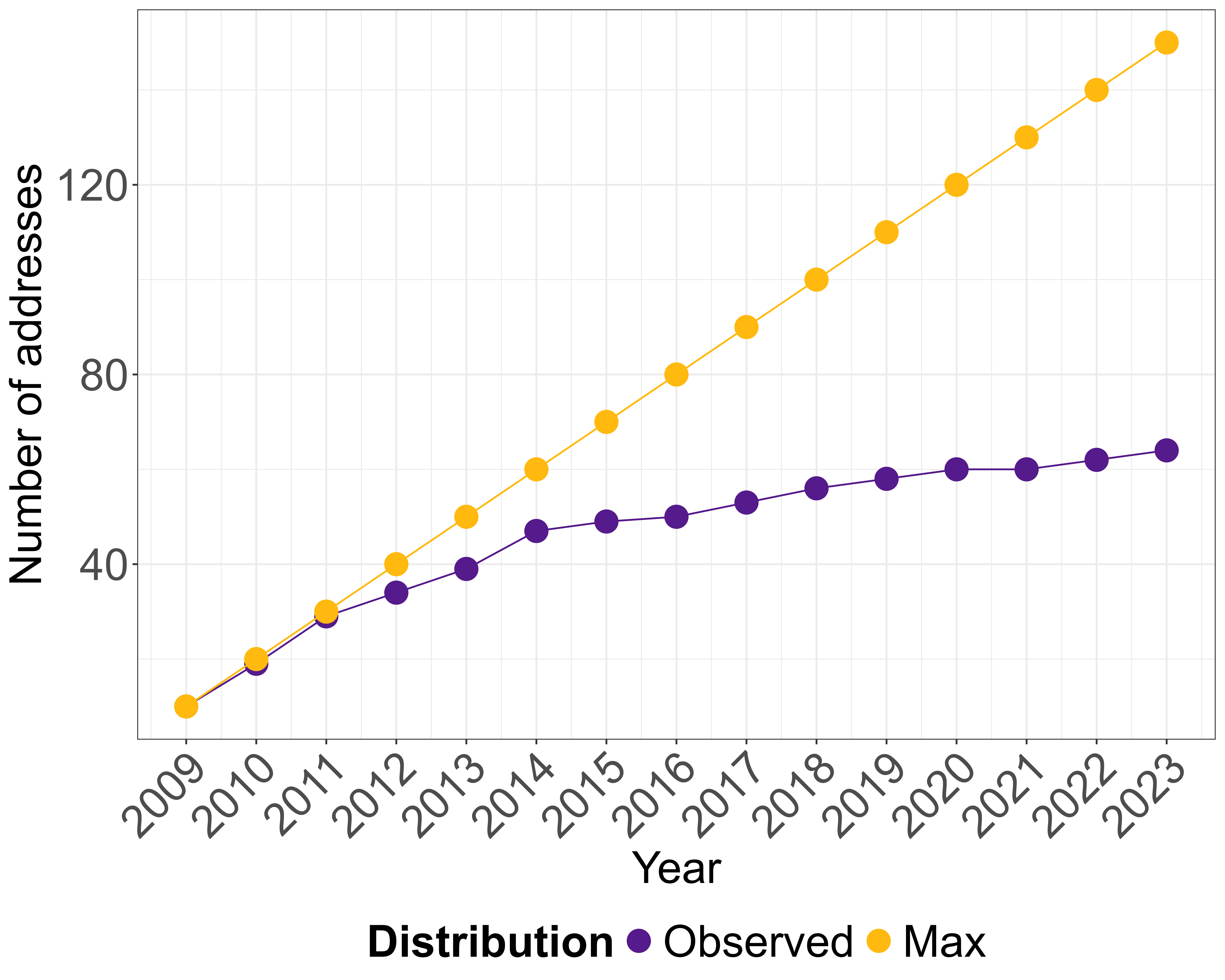}%
\label{figure7a}}
\hfil
\subfloat[]{\includegraphics[width=2.5in]{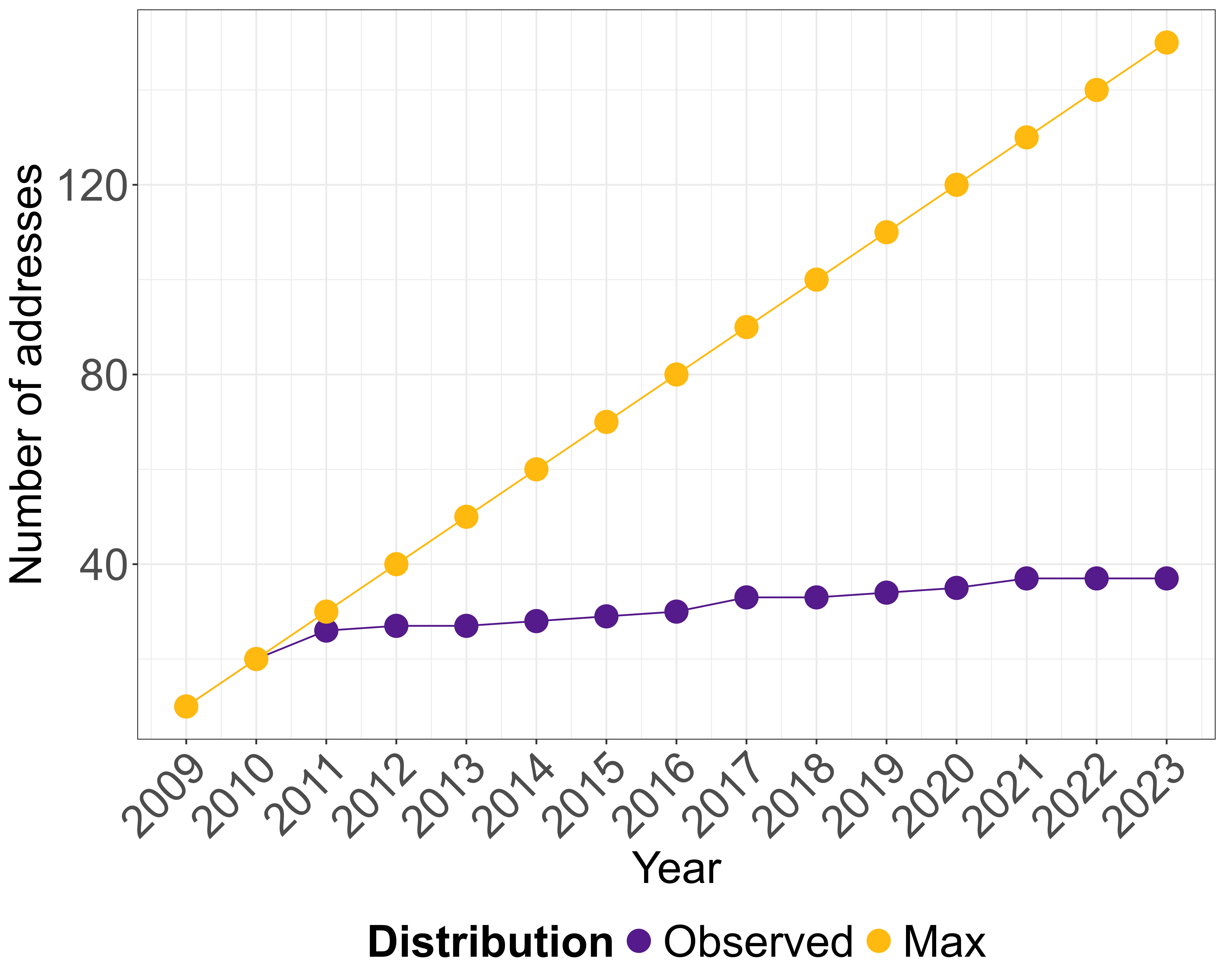}%
\label{figure7b}}
\caption{Evolution of $X_{b,t}$ (a) and $Y_{i,t}$ (b) and maximum possible values over time; $X_{b,t}$ and $Y_{i,t}$ are defined as in \hyperref[formula7]{Equation 7}.}\label{figure7}
\end{figure*}

\textit{Hypothesis 2} can be tested by looking at the evolution of the union set of the richest nodes over the 15 snapshots, \hyperref[figure7]{Figure 7} illustrates this trend over time. The curves represent, respectively, the theoretical maximum dimensions of the union set of the $k$ nodes and the observed evolution of these quantities. The curves of the observed values remained clearly below the expected maximum lines. In particular, the stability in the control of in-degree activity exceeds that of wealth accumulation; this could be explained by the fact that it is easier to accommodate more incoming transactions than to significantly increase the total wealth accumulated, so that a node can receive numerous transactions of low value and be among the top $k$ nodes only for the in-degree statistics. 

This also led us to confirm \textit{Hypothesis 2}: the richest nodes tend to maintain their status throughout the period, they do not lose their top ten placement over time. Note that we were unable to de-anonymise some of the nodes from the early years due to insufficient information, but this actually strengthens our results, as we have already captured all possible address heterogeneity in the early years.

\begin{figure*}[!t]
\centering
\subfloat[]{\includegraphics[width=2.5in]{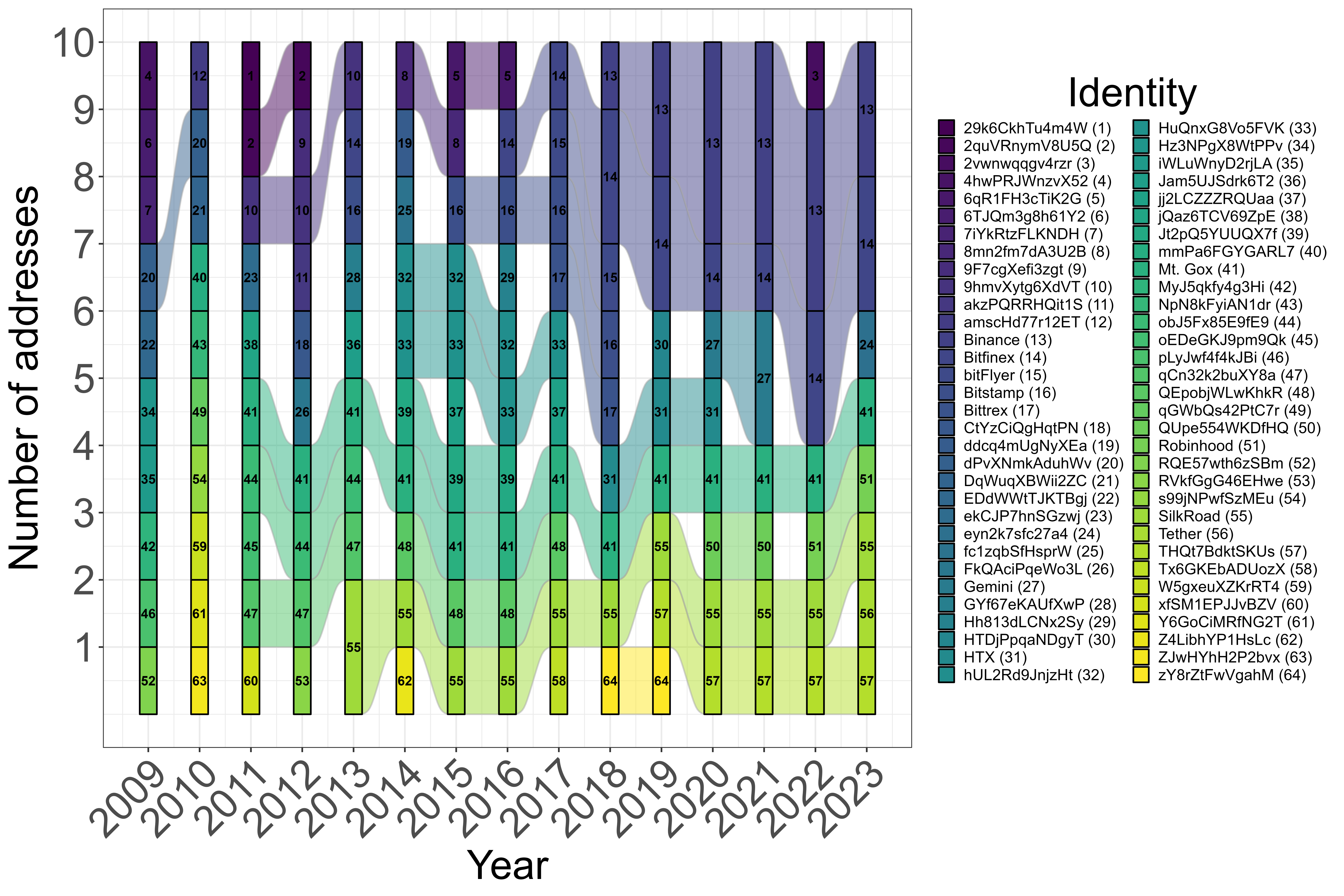}%
\label{figure8a}}
\hfil
\subfloat[]{\includegraphics[width=2.5in]{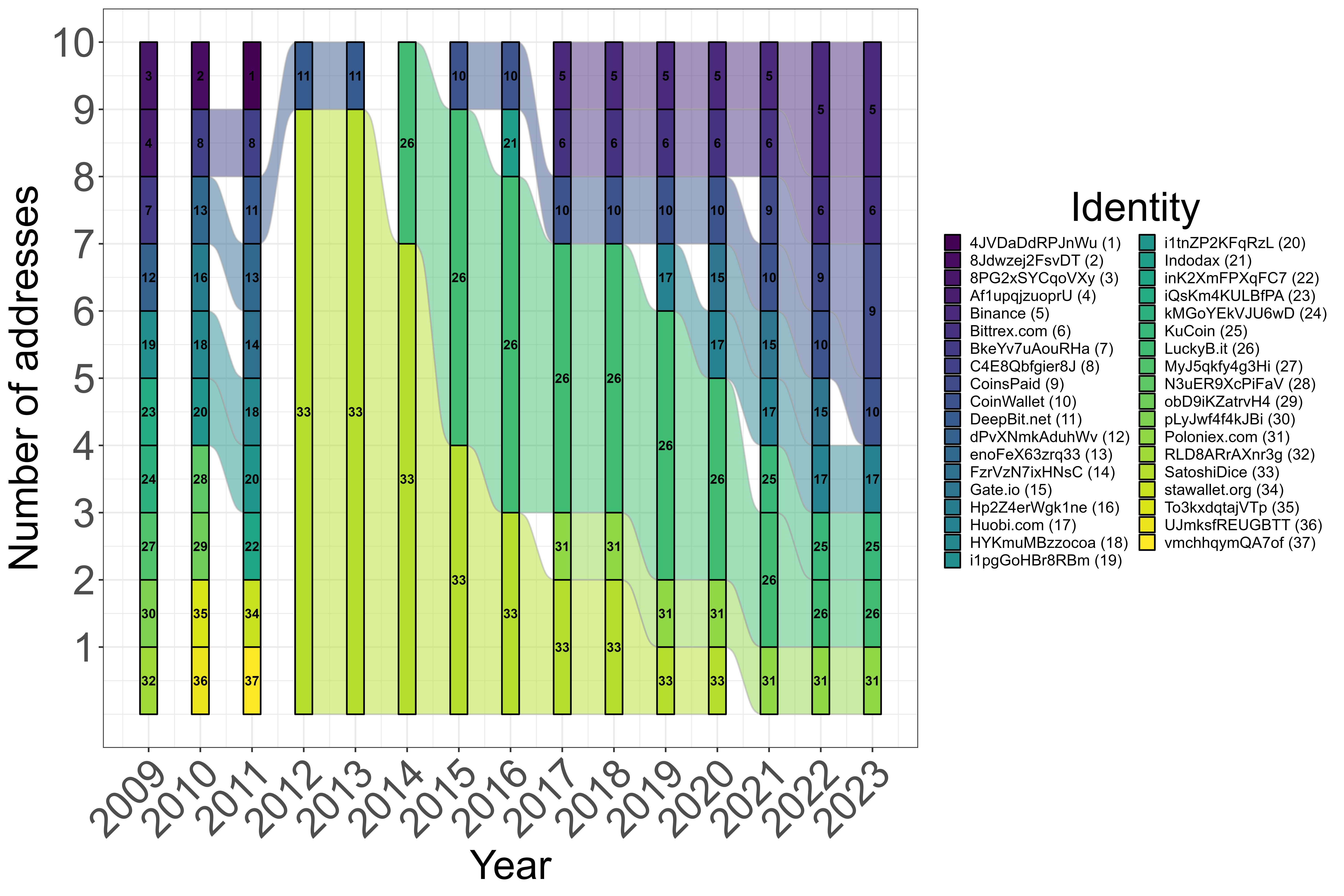}%
\label{figure8b}}
\caption{Evolution of the set of the richest nodes over time, by balance and in-degree.}\label{figure8}
\end{figure*}

\hyperref[figure8]{Figure 8} confirms our findings on \textit{Hypothesis 2}. \hyperref[figure8]{Figures 8a} and \hyperref[figure8]{8b} show, respectively, the evolution of the sets of the richest addresses in terms of balance and in-degree. In particular, both figures show the persistence of the same addresses among the richest, e.g. a wallet linked to Mt. Gox (see \hyperref[figure8]{8a}) has been in the top 10 since 2011. Moreover, the nodes that emerge as the richest are increasingly exchanges that act as intermediaries in the system. Finally, as noted for \hyperref[figure7]{Figure 7}, we observe less variation in terms of in-degree richness than in terms of balance as it is much easier to receive incoming transactions than to accumulate wealth.

Overall, these results provide clear support for the rich-get-richer mechanism found by \cite{Maesa_al_2018} up to 2015. Furthermore, we argue that this mechanism contributed to the creation of path dependencies in the system: once a node enters the richest group, it tends to stay in it and continue to accumulate more resources. It is important to highlight that these dependencies persist even after the entry of institutional actors, as the composition of the union sets had already started to stabilise by 2014, before any significant institutional involvement.

\section{Discussion and Conclusions}\label{discussion}

The creation of Bitcoin marked a turning point in the evolution of the financial systems and institutions. Characterised by decentralisation, trustless transactions, and an alleged egalitarianism \cite{corradi2021too}, the evolution of Bitcoin has revealed a puzzling path: centralisation tendencies have become apparent, as noted by \cite{ron2013quantitative} and \cite{kondor2014rich}, systemic bottlenecks have emerged and Bitcoin has started to be populated by institutional actors and retailers, as well as being  subject to increasing regulation.

In our study, we wanted to explore the unplanned evolution of Bitcoin's network structures and dynamics, focusing on centralisation trends, network characteristics and structures, and testing the rich-get-richer mechanism. We first characterised the Bitcoin transaction network by mapping its structures and dynamics over fifteen one-year snapshots. The longitudinal empirical approach then allowed us to study evolutionary trends from the system's inception to its most recent developments. We performed a network analysis that revealed persistently low density values -typical of large-scale networks- similar to what \cite{nerurkar2021dissecting} found in their study. 

Furthermore, in line with most of the previous research, we found right-skewed distributions of nodes' in-degree and out-degree indicating a pronounced centralisation of node activity and wealth accumulation as in \cite{kondor2014rich} and \cite{Maesa_al_2018, The_Anti-Social}, with a small minority of players dominating the network. In particular, we focused on centralisation and concentration trends, which are contrary to expectations for a system like Bitcoin. As in the previous studies by \cite{reid2013analysis, kondor2014rich, baumann2014exploring, The_Anti-Social}, we found skewed distributions and high levels of inequality from the very first years of the system's life, since what we defined as the \textit{Exploration} phase, a period characterised by early adopters and low levels of activity. Therefore, we excluded any role of external regulation and institutional actors in the emergence of the centralisation trend: it emerged endogenously from the free interactions between addresses in the network. Our results showed that these distributions were highly unequal, with Gini indices ranging from 0.75 to 1, in line with earlier studies in the field \cite{The_Anti-Social, kondor2014rich, Maesa_al_2018}.  

We also found evidence for the rich-get-richer mechanism, which we investigated through the accumulation of nodes' wealth and activity and its persistence over time. While this was  already found for the first years of Bitcoin activity by \cite{Maesa_al_2018}, our results confirm that this pattern has even strengthened over time. This path-dependent behaviour is indicative of a system in which initial advantages are reinforced over time, shaping the emergent properties of the network. Once an early node becomes one of the richest, it is rarely displaced. Therefore, our analysis revealed the early onset of path dependencies behind these network trends. 
The immutable nature of the blockchain -once a block is added it cannot be removed or modified- may facilitate this path dependency, while the limited and diminishing supply of new tokens may have encouraged wealth accumulation among early network participants. However, this should not be taken for granted; the growing role of financial institutions, defined by \cite{baldwin2018digital} as the corporatisation of Bitcoin, should not be neglected. This is confirmed by the small but significant changes in the set of the richest nodes (see \hyperref[methods.rich]{Section 4.2}): over time, the richest nodes are increasingly institutions rather than anonymous owners or early adopters.

We found persistent disassortative mixing tendencies, thus confirming previous evidence by \cite{kondor2014rich, The_Anti-Social,motamed2019quantitative, WU_al_2021}: although small, these tendencies are indicative of a system in which low-connected nodes preferentially connect with high-connected node. This pattern suggests a preference for trading with one's opposite in terms of network activity. Furthermore, while a positive transitivity coefficient is not new in Bitcoin \cite{kondor2014rich}, we found a slightly stronger clustering tendency compared to previous results  \cite{nerurkar2021dissecting}. The clustering, although low, warrants careful consideration in future studies, as it may imply a variety of occurring triadic configurations relevant for understanding and explaining the network evolution.

Focusing on the network components, we found that almost the entire network is indirectly connected, with no relevant disconnected components. This confirms the findings of \cite{The_Anti-Social} and \cite{nerurkar2021dissecting}, who also found a giant component where the majority of nodes were at least indirectly connected. We also found a large strongly connected component that emerged as a focal point in the network; extending \cite{The_Anti-Social, nerurkar2021dissecting}, we observed a higher proportion of out-degree rich addresses than in-degree rich nodes in it, as the former are more important for keeping this network component fully connected.

Finally, by studying the dynamics of our measures and the history of Bitcoin, we defined three periods in the evolution of Bitcoin: the \textit{Exploration} phase, the \textit{Adaptation} phase and the \textit{Maturity} phase. The exploration phase was characterised by  early adopters and low activity levels until the first halving (2012), a period in which Bitcoin incubated its disruptive evolution; the adaptation phase saw the shutdown of Silkroad -one of the most famous dark markets-, the collapse of Mt. Gox (2014) -the largest exchange at the time- and the first regulatory attempts by the US Internal Revenue Service. The maturity phase then began in 2015 and was characterised by decreasing volatility, increasing activity and participation, and greater institutional involvement. During this phase, not only was Bitcoin less stigmatised than before, but we also observed a stabilisation in our network measures after 2015, meaning that the Bitcoin internal functioning and its external evolution are linked. In addition to \cite{kondor2014inferring} and \cite{tasca2018evolution}, we mention that considering these two aspects together -metric dynamics and Bitcoin-related events- is key to capture the complexity of network systems such as Bitcoin and cryptocurrencies in general.

In summary, our study has made three main contributions. First, we used a new, comprehensive dataset that extends previous studies that only considered small time windows for the analysis. This allowed us to provide an updated perspective on the network's evolution, which could inspire future network research at even finer temporal scales. Second, we provided updated results on the centralisation and concentration patterns of a decentralised cryptocurrency system, and found evidence for the strength of the rich-get-richer mechanism. Third, by combining the dynamics of the observed metrics and the Bitcoin-related events, we identified three periods in the evolution of Bitcoin that better account for endogenous and exogenous events.

However, the broad scope of Bitcoin and the extensive information provided by the blockchain cannot be captured in a single study. The timestamped nature of Bitcoin was not fully explored in our study and future studies should focus on this aspect with more fine-grained analysis: for example, applying an event history analysis or a relational event model (e.g., \cite{LOMI2021, bianchi_lomi2023, bianchi2024relational}) to Bitcoin transaction data would enrich our understanding of the endogenous and exogenous forces beyond the evolution of this system, but this would require considering a higher granularity of timestamps. Using shorter time intervals would also help us to better account for significant fluctuations in network dynamics at a finer resolution.  This would allow, for example, to focus on specific periods of extremely high price volatility in order to study network dynamics in such a context, which may show peculiar behaviour. This is because a longer observation period, as we considered in our study, may capture long-term trends, but miss the specificity of different events that may have an impact on the Bitcoin network.

In conclusion, our study shows the relevance of studying Bitcoin as a complex system, highlighting its dynamics and network structures. The availability of fine-grained microdata on financial transactions offered by the blockchain is unique, but has not yet been sufficiently explored. In this context, we proposed a study that characterised the complexity of Bitcoin through its network of addresses and transactions, thus considering the micro-level interactions of the system. Observing the rise in the price of bitcoins in November 2024 and the general expectations about the future of cryptocurrencies by political leaders, companies and regulators, we need to mention once again the relevance of studying cryptocurrencies with computational network research. Monitoring the evolution of cryptocurrencies at different levels of analysis, and hopefully with more comparative approaches, can help to understand whether these new economic infrastructures really generate new properties and can be resilient to external shocks compared to traditional money markets. Here, in line with previous research \cite{kondor2014rich, Maesa_al_2018, nerurkar2021dissecting}, our results cast some doubt on at least the former, as centrality, inequality and power concentration seem to be similar to traditional market infrastructures.

\section{Data Availability}
Data from the Bitcoin blockchain is available in Google BigQuery. Refer to the bigquery-public-data.crypto\_bitcoin dataset.
Data for replicating the outcomes of this manuscript is available at \url{https://doi.org/10.13130/RD_UNIMI/NFURUT}.

\section{Funding}
This research did not receive any specific grant from funding agencies in the public, commercial, or not-for-profit sectors.

\section{Author contributions}
MV coordinated and designed the study, designed and executed the analysis, wrote and revised the manuscript, DG-C built the dataset, executed the analysis and revised the manuscript, EA-G built the dataset, executed the analysis and revised the manuscript, FG coordinated the data collection, executed the analysis and revised the manuscript. FS designed the study, revised the analysis, wrote and revised the manuscript.

\section{Competing interests}

The authors declare no competing interests.

\section{Acknowledgements}
We would like to thank Carmen Guarner for her preliminary work on data extraction and pre-processing and for her help with the first steps of our study.

\bibliographystyle{IEEEtran} 
\bibliography{bibliography}

\begin{IEEEbiographynophoto}{Marco Venturini} (corresponding author: marco.venturini@unimi.it)
is a PhD Student in Economic Sociology at the Department of Social and Political Sciences, University of Milan, Milan, Italy and at the Faculté des Lettres, Sorbonne Université, Paris, France. He is also member of Behave Lab (Unimi) and GEMASS (SU). 
\end{IEEEbiographynophoto}
\begin{IEEEbiographynophoto}{Daniel García-Costa}
is an Assistant Professor and member of the LAIA-UV (Learning, Artificial Intelligence and Applications) research group at the Department of Computer Science, University of Valencia, Burjassot, Spain.
\end{IEEEbiographynophoto}
\begin{IEEEbiographynophoto}{Elena Álvarez-García}
is a PhD Student and member of the LAIA-UV (Learning, Artificial Intelligence and Applications) research group at the Department of Computer Science, University of Valencia, Burjassot, Spain.
\end{IEEEbiographynophoto}
\begin{IEEEbiographynophoto}{Francisco Grimaldo}
is a Full Professor of Computer Science and Artificial Intelligence and head of the LAIA-UV research group at the Department of Computer Science, University of Valencia, Burjassot, Spain. He is also Director of the Capgemini-UV Chair in innovation in software development.
\end{IEEEbiographynophoto}
\begin{IEEEbiographynophoto}{Flaminio Squazzoni}
is a Full Professor of Sociology and head of the Behave Lab at Department of Social and Political Sciences, University of Milan, Milan, Italy.
\end{IEEEbiographynophoto}

\end{document}